\documentclass[twocolumn,floatfix,prb,aps,showpacs]{revtex4}
\usepackage{graphicx,amsmath,amssymb}

\begin{document}

\title{Particle-hole symmetry and bifurcating ground state manifold in the quantum Hall ferromagnetic states
of multilayer graphene}

\author{Csaba T\H oke}
\affiliation{BME-MTA Exotic Quantum Phases ``Lend\"ulet'' Research Group, Budapest University of Technology and Economics, Institute of Physics, Budafoki \'ut 8., H-1111 Budapest, Hungary}
\date{\today}

\begin{abstract}
The orbital structure of the quantum Hall ferromagnetic states in the zero-energy Landau level
in chiral multilayer graphene (AB, ABC, ABCA, etc.\ stackings)
is determined by the exchange interaction with all levels, including deep-lying states in the Dirac sea.
This exchange field favors orbitally coherent states with a U(1) orbital symmetry if
the filling factor $\nu$ is not a multiple of the number of layers.
If electrons fill the orbital sector of a fixed spin/valley component to one-half, e.g.,
at $\nu=\pm3,\pm1$ in the bilayer and at $\nu=\pm2,\pm6$ in the ABCA four-layer,
there is a transition to an $Z_2\times$U(1) manifold.
For weak interaction, the structure in the zero-energy Landau band compensates for the
different exchange interaction on the sublattices in the Landau orbitals;
on the other side, the ground state comes in two copies that distribute charge on
the sublattices differently.
We expect a sequence of similar bifurcations in multilayers of Bernal stacking.
\end{abstract}


\maketitle

\newcommand{\bsigma}{\mbox{\boldmath$\sigma$}}

In their pristine form, monolayer graphene,\cite{graphene} bilayer graphene\cite{bilayer} and trilayer graphene\cite{trilayer} are zero-gap semiconductors with
a Landau level at zero energy, i.e., where the valence and the conduction bands touch in the absence of a magnetic field.
This level significantly affects the integer quantum Hall effect\cite{Klitzing} (IQHE).
Quantum Hall ferromagnetism (QHF),\cite{qhf} combined with smaller terms such as the Zeeman energy, fully resolves this level,\cite{qhfmonolayer,qhfbilayer}
Particularly interesting are the multilayers, because their zero-energy Landau band (ZLB) has orbital degeneracies above the ubiquitous spin and valley quasidegeneracies.

Recently, Shizuya has pointed out that the exchange field created by the sea of filled Landau levels (LL's)
is essential for correctly identifying the QHF ground states in bilayer graphene\cite{Shizuya1}
and ABC trilayer.\cite{Shizuya2}
This exchange field favors orbitally coherent states, in contrast to the Hund's rule picture\cite{Hund,phsymmbreaking}
that predicts the order the zero-energy orbitals are filled.
We make use of this observation, but no longer treat the Dirac sea as inert.
We identify a number of QHF states, and find that the interplay of the sublattice structure and exchange
leads to bifurcations of the ground state manifold if the number of layers $s$ is even
and the filling factor is such as to half-fill the orbital sector of fixed spin and valley at zero energy.

\textit{Hamiltonians.}
The following class of Hamiltonians concisely describes the low-energy physics of chiral (rhombohedral) stacking series, i.e., the AB, ABC, ABCA... multlilayers ($s>1$ is the number of layers)
if the layers are energetically equivalent,
\begin{equation}
\hat H^{(s)}_\xi=\frac{v^2}{(-\gamma_1)^{s-1}}
\begin{pmatrix}
0 & (\pi^\dag)^s \\ \pi^s & 0
\end{pmatrix}
-g\mu_B \mathbf B\cdot\bsigma,
\end{equation}
where $\pi=p_x+ip_y$ with $\mathbf p=-i\hbar\nabla -e\mathbf A$,
$v=\sqrt3 a\gamma_0/2\hbar\approx 10^6$ m/s is derived from the $\gamma_0$ intralayer nearest-neighbor hopping amplitude,
and $\gamma_1\approx0.4$ eV is the interlayer hopping between dimer sites (i.e., sites exactly above/below one another).

For bilayer graphene, $\hat H^{(2)}_\xi$ is obtained by a Schrieffer-Wolf transformation\cite{tightbinding} from the
Slonczewski-Weiss-McClure tight-binding model of graphite,\cite{SWM}
expanded to first order in the momentum difference $\mathbf q-\xi\mathbf K$ from the nonequivalent corners $\mathbf K,-\mathbf K$ of the hexagonal first Brillouin zone.
This Hamiltonian is block diagonal in the valley index; it acts on spinors built up of the amplitudes $[\psi_{A1},\psi_{B2}]$ in valley $\mathbf K$ and
$[\psi_{B2},\psi_{A1}]$ in valley $-\mathbf K$, where $A2$ and $B1$ are dimer sites, while the $A1$ and $B2$ sites (nondimer sites)
are above/below the center of a hexagon in the other layer.
The interlayer next-nearest neighbor hoppings $\gamma_3=\gamma_{A1,B2}$ and $\gamma_4=\gamma_{A1,A2}=\gamma_{B1,B2}$ have been neglected,
just like the on-site energy difference $\Delta'$ between the dimer sites ($A2,B1$) and the nondimer sites ($B2,A1$).

If the sublattices are denoted $An$ and $Bn$, so that $(B1,A2)$ and $(B2,A3)$ are dimer site pairs for the ABC trilayer graphene,
$\hat H_\xi^{(3)}$ acts on amplitudes $(A1,B3)$ in valley $\mathbf K$ and $(B3,A1)$ in valley $-\mathbf K$.\cite{Koshino}
The second layer hopping $\gamma_2\equiv\gamma_{A1,B3}$ has also been neglected.
For four-layers in the ABCA stacking ($s=4$), the approximations are the same.

Hamiltonians $\hat H^{(s)}_\xi$ have both negative and positive energy LL's at
$\epsilon^{(s)}_n=\text{sgn}(n)\varepsilon_s\sqrt{\prod_{i=0}^{s-1}(|n|-i)}$,
with $\varepsilon_s\equiv(2\hbar v^2eB)^{s/2}/\gamma_1^{s-1}$, and $s$ zero energy LL's, $0\le n<s$.
The orbitals are
$\Psi_{|n|\ge s,q\xi}(\mathbf r)=\frac{1}{\sqrt2}\begin{pmatrix} \eta_{|n|q}(\mathbf r) \\ \text{sgn}(n)\eta_{|n|-s,q}(\mathbf r)\end{pmatrix}$ and
$\Psi_{0\le n<s;q\xi}(\mathbf r)=\begin{pmatrix} \eta_{nq}(\mathbf r) \\ 0\end{pmatrix}$,
where $\eta_{nq}(\mathbf r)=\frac{e^{iqx-\left(y/\ell-q\ell\right)^2/2}}{\sqrt{2\pi\sqrt\pi 2^n n!\ell}}H_n\left( \frac{y}{\ell}-q\ell \right)$
are the single-particle states in the conventional two-dimensional electron gas with guiding center position at $y=q\ell^2$
in the Landau gauge $\mathbf A=\mathbf{\hat y}Bx$.
$\ell=\sqrt{\hbar/(eB)}$ is the magnetic length, and the filling factor is $\nu=2\pi\ell^2\rho$.

Suppressing spin and valley indices to avoid clutter, the long-range part of the Coulomb interaction is
\begin{multline}
\label{coulomb}
\hat V=\frac{e^2}{8\pi\epsilon}
\sum_{n_1n_2n_3n_4}\sum_{p_1p_2}\int\frac{d^2k}{(2\pi)^2}
F^{(s)}_{n_1n_4}(\mathbf k)F^{(s)}_{n_2n_3}(-\mathbf k)\frac{2\pi}{k}\\
\hat c^\dag_{n_1p_1}\hat c^\dag_{n_2p_2}\hat c_{n_3,p_2-k_y}\hat c_{n_4,p_1+k_y}e^{ik_x\ell^2(p_2-p_1-k_y)},
\end{multline}
where $\hat c_{np}$ annihilates an electron with quantum numbers $n$ and $p$,
and the form factor $F^{(s)}_{nn'}(\mathbf k)$ is
$F^{(s)}_{nn'}(\mathbf k)=F_{nn'}(\mathbf k)$ if $0\le n,n'<s$;
$\frac{1}{\sqrt 2}F_{|n||n'|}(\mathbf k)$ if either $0\le n<s$ or $0\le n'<s$; and
$\left(F_{|n||n'|}(\mathbf k)+\text{sgn}(nn')F_{|n|-s,|n'|-s}(\mathbf k)\right)/2$ if $|n|,|n'|\ge s$;
otherwise it is undefined.
We have used
\begin{equation*}
F_{n'n}(\mathbf k)=\sqrt\frac{n!}{(n')!}\left(\ell\frac{k_y-ik_x}{\sqrt2}\right)^{n'-n}L_n^{n'-n}\left(\frac{k^2\ell^2}{2}\right)e^{-k^2\ell^2/4},
\end{equation*}
if $n'\ge n$, else $F_{nn'}(\mathbf k)=F^\ast_{n'n}(-\mathbf k)$.
We have neglected the difference between the intralayer and interlayer interactions, because the distance $d=0.335$ nm between the layers is much less than $\ell$.
The electron-electron interaction also includes a short-range part that reflects the symmetry of the Bravais lattice.
This part determines the spin and valley structure of several QHF states,\cite{Kharitonov}
but it is not of interest in our study.
Let us characterize the relative strength of the interaction by $\beta^{(s)}=e^2/4\pi\epsilon\ell\varepsilon_s\propto B^{(1-s)/2}$.

For our purposes, the minimal models $\hat H^{(s)}_\xi$
are applicable if the splitting of the ZLB due to the neglected hoppings such as the trigonal warping term $\gamma_3$
is small (see Supplementary Material),\cite{supp} and the moderate energy orbitals are only slightly distorted by the closeness of the Lifshitz
transition saddle point \cite{tightbinding,Koshino} on the low end and
the split bands around $\pm\gamma_1$ at the high end.\cite{supp,Sari}
The latter requires $\beta^{(2)}\lesssim 17/(\epsilon_r\sqrt{B\text{ [T]}})\approx9.2$,
$\beta^{(3)}\lesssim 188/(\epsilon_r B\text{ [T]})\approx38$, and
$\beta^{(4)}\lesssim 2073/(\epsilon_r B^{3/2}\text{ [T]})\approx41$, where we have estimated
$\epsilon_r\approx4$ in the final numbers;\cite{supp}
and for $n\lesssim 120/B[\text{T}]$ the LLs are far from $\pm\gamma_1$ for any $s$.\cite{supp}
The two-band model slightly overestimates the LL energies;\cite{Cote} these ensuing quantitative modifications
will be discussed below.

We do standard mean-field decomposition of $\hat V$.
The direct (Hartree) term $\hat V^\text{H}$ is canceled as usual by the background for any homogeneous state.
The exchange (Fock) term $\hat V^\text{F}$ can be written as
\begin{multline}
\label{meanfield}
\frac{\hat V^\text{F}}{N_\phi}=-\frac{e^2}{4\pi\epsilon}\sum_{nn'mm'}\sum_{\xi\xi'\sigma\sigma'}
\int\frac{d^2k}{(2\pi)^2}F^{(s)}_{nm}(\mathbf k)F^{(s)}_{m'n'}(-\mathbf k)\frac{2\pi}{k}\\
\left[\left\langle \hat c^\dag_{mp\xi\sigma}\hat c_{m'p\xi'\sigma'} \right\rangle \hat c^\dag_{n,p-k_y,\xi'\sigma'}\hat c_{n'_,p-k_y,\xi\sigma}-\right.\\
\left.-\frac{1}{2}\left\langle\hat c^\dag_{mp\xi\sigma}\hat c_{m'p\xi'\sigma'}\right\rangle\left\langle \hat c^\dag_{np-k_y,\xi'\sigma'}\hat c_{n',p-k_y,\xi\sigma}\right\rangle\right],
\end{multline}
where $p$ is arbitrary but fixed and $N_\phi$ is the number of flux quanta piercing the sample.
The angular integral vanishes unless $|n|-|m|=|n'|-|m'|$.

\begin{figure}[htbp]
\begin{center}
\includegraphics[width=\columnwidth,keepaspectratio]{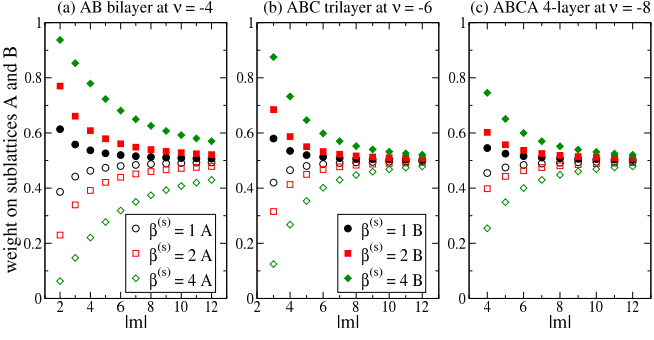}
\end{center}
\caption{\label{iqhe}(Color online)
In the integer quantum Hall effect at $\nu=-2s$ of chiral $s$-layer graphene,
the mixing of orbitals of index $\pm m$ produces an unequal charge distribution
$w^\pm(\theta_m,\phi_m)=(1\pm\sin\theta_m\cos\phi_m)/2$ on the
sublattices that belong to the top spinorial component (empty symbols) and the bottom one (filled symbols).
At $\nu=2s$ the sublattices are interchanged.
}
\end{figure}

At $\nu=-2s$ the ZLB is empty; the single-particle gap gives rise to an IQHE.
For rotationally invariant states, the exchange interaction $\hat V^\text{F}$ mixes LL's $m,m'$ only
if $|m|=|m'|$.\cite{Barlas3,supp}
Seeking the ground state in the parametric form
$\Psi^{(s)}_{0}=\prod_{m\le-s}\prod_{q,\xi,\sigma}
\left(\cos\left(\theta_{m\xi\sigma}/2\right)e^{i\phi_{m\xi\sigma}/2}\hat c^\dag_{mq\xi\sigma}\right.$+
$\left.\sin\left(\theta_{m\xi\sigma}/2\right)e^{-i\phi_{m\xi\sigma}/2}\hat c^\dag_{-m,q\xi\sigma}\right)\left|0\right\rangle,$
the ground-state energy is (suppressing $\xi,\sigma$ for brevity)
\begin{gather}
\label{zerothenergy}
\frac{E^{(s)}_0}{N_\phi}=
-\varepsilon_s\sum_{m=s}^\infty\sqrt{\prod_{i=0}^{s-1}(m-i)}\cos\theta_{m}-\\
-\frac{1}{2}\sum_{n=s}^\infty\sum_{m=s}^\infty
\left[X_{nm}w^+(\theta_m,\phi_m)w^+(\theta_n,\phi_n)\right.+\nonumber\\
+X_{n-s,m-s}w^-(\theta_m,\phi_m)w^-(\theta_n,\phi_n)+\nonumber\\
\left.+\frac{1}{2}X^{(s)}_{nm}\left(\cos\theta_m\cos\theta_n-\sin\theta_m\sin\phi_m\sin\theta_n\sin\phi_n\right)\right],
\nonumber
\end{gather}
where we have defined $w^\pm(\theta,\phi)=(1\pm\sin\theta\cos\phi)/2$,
$X_{n_1n_2n_3n_4}=\frac{e^2}{4\pi\epsilon}\int\frac{d^2k}{(2\pi)^2}F_{n_1n_2}(\mathbf k)F^\ast_{n_3n_4}(\mathbf k)\frac{2\pi}{k}$,
$X_{nm}=X_{nmnm}$, and $X^{(s)}_{nm}=X_{nm,n-s,m-s}$.\cite{supp}
[The ratio of the first term in Eq.~(\ref{zerothenergy}) to the subsequent terms depends on $\beta^{(s)}$.]
Here $w^\pm(\theta_m,\phi_m)$ is the weight the electron in the fixed $|m|$ subspace has in the sublattice that corresponds to the top (bottom) spinor component.
For minimizing $E^{(s)}_0$ numerically, an LL cutoff $-M$ is introduced.
We set $\theta_{m>M,\xi\sigma}=0$ and utilize the identity $\sum_{n=0}^\infty|F_{n'n}(\mathbf k)|^2=1$ for arbitrary $n'$,
for reducing infinite summations of exchange integrals to finite sums,
plus a featureless divergent constant we omit.\cite{Shizuya1,Shizuya2,supp}
The exchange field is still due to \textit{all} filled levels; we merely treat the low-lying ones as inert.
The charge distribution $w^\pm$ for the optimized parameters are shown in Fig.~\ref{iqhe} for $M=12$; larger cutoffs yield practically identical results.
Only the topmost LL's are affected by the interaction.
As the exchange between lower-$n$ spinor components $\eta_{nq}$ is larger, the filled state in the fixed $|m|$ subspace is
selected so that it increases its weight in the lower component.
This tendency is limited by the kinetic energy.
While the charge transfer between sublattices is infinitesimal in the infinite band width model we consider,
it is finite in a lattice description, where the direct (Hartree) interaction also counteracts it.

The completely filled ZLB at $\nu=2s$ is treated by the Ansatz
$\Psi^{(s)}_{s}=\prod_q\prod_{n=0}^{s-1}\hat c^\dag_{nq}\Psi^{(s)}_0$.
The optimal states $\Psi^{(s)}_0$ and $\Psi^{(s)}_s$ are related:\cite{supp}
if $\{\theta_m,\phi_m\}$ optimizes $\Psi^{(s)}_0$, $\{\theta_m,\phi_m+\pi\}$ optimizes $\Psi^{(s)}_s$.
The preferential sublattices are interchanged.
This is a manifestation of particle-hole symmetry.

At $\overline\nu = \nu+4=1$ relative filling of the ZLB of \textit{bilayer graphene},
all of the spin, valley, and orbital symmetries must be broken.
Assume the first two have already been broken, as dictated by the 
tiny but nonzero Zeeman energy $g\mu_BB\approx1.3 B[\text{T}]$ K,
and, as valley is equivalent to layer in the ZLB,\cite{tightbinding}
by the minimization of the capacitive energy.\cite{supp}
We seek the translation-invariant ground state as
$\Psi^{(2)}_{1}=\prod_q\left(z_0\hat c^\dag_{0q}+z_1\hat c^\dag_{1q}\right)\Psi^{(2)}_0$, which has the energy,
\begin{multline}
\label{bilayeronegeneral}
\frac{E^{(2)}_{1}}{N_\phi}=\frac{E^{(2)}_0}{N_\phi}
-\frac{X_{00}}{2}|z_0|^4-\frac{X_{11}}{2}|z_1|^4-(X_{01}+X_{0011})|z_0|^2|z_1|^2-\\
-\sum_{m=2}^\infty\left(|z_0|^2 X_{0m}+|z_1|^2 X_{1m}\right)w^+(\theta_m,\phi_m).
\end{multline}

The weak-coupling solution ($\beta^{(2)}\to0$) is $|z_0|^2=|z_1|^2=\frac{1}{2}$ and the relative phase of
$z_0,z_1$ is irrelevant;
the state has orbital coherence (possibly above layer coherence), and it is its own particle-hole conjugate.
The exchange field of the filled Dirac sea acts differently on the $n=0,1$ orbitals,
in analogy to the Lamb shift in QED.\cite{Shizuya1}
This in turn removes the advantage of placing all electrons in the $n=0$ orbitals, which would minimize the
intra-ZLB part of the exchange interaction.\cite{Hund}

\begin{figure}[htbp]
\begin{center}
\includegraphics[width=\columnwidth,keepaspectratio]{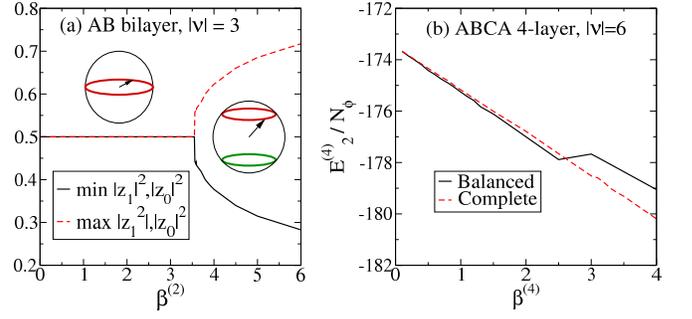}
\end{center}
\caption{\label{params1}(Color online)
Bifurcation of the ground-state manifold.
(a) In the bilayer graphene at $\overline\nu=1$, the orbitally coherent ground state
comes in two symmetry-related copies if the interaction is strong enough, $\beta^{(2)}>\beta^{(2)}_\text{c}\approx3.56$.
The insets show the circle(s) the orbital isospin $z_0\equiv\cos\frac{\Theta}{2}e^{i\Phi/2}$,
$z_1\equiv\sin\frac{\Theta}{2}e^{-i\Phi/2}$ maps out in the two regions.
(b) In the ABCA four-layer at $\overline\nu=2$ the ground state no longer
belongs to the balanced subspace for $\beta^{(4)}>\beta^{(4)}_\text{c}\approx2.6$;
thus it must come in symmetry-related pairs.
}
\end{figure}

At finite $\beta^{(2)}$ we minimize $E^{(2)}_{1}$ with a LL cutoff $-M$, keeping the $m<-M$ LLs but treating them as inert.
Remarkably, the above structure of the ZLB is preserved for finite interaction strength up to
$\beta^{(2)}_\text{c}\approx3.56$ [see Fig.~\ref{params1}(a)].
For this state $\theta_m=0$.
Physically, with an equal weight of $n=0,1$ orbitals filled there is no preferential sublattice, and
the optimization in the $|m|\ge2$ subspaces decouples from the rest.
For $\beta^{(2)}>\beta^{(2)}_\text{c}$, the gain of exchange energy from an unequal filling of sublattices compensates
the kinetic energy cost; such states must come in pairs,
as $z_0\leftrightarrow z_1$ and $\phi_m\leftrightarrow\phi_m+\pi$ is an exact symmetry of Eq.~(\ref{bilayeronegeneral}).
The symmetry of the ground-state manifold increases from U(1) to $Z_2\times$U(1) at $\beta^{(2)}=\beta^{(2)}_\text{c}$.
The coefficients of the ZLB structure can be identified with an ``isospin''.
For $\beta^{(2)}<\beta^{(2)}_\text{c}$ this points to the equator of the Bloch sphere, while 
for $\beta^{(2)}>\beta^{(2)}_\text{c}$ it maps out two parallel circles in the two hemispheres
[insets of Fig.~\ref{params1}(a)].
$\beta^{(2)}_\text{c}=3.56$ corresponds to $B_\text{c}=23\text{ T}/\epsilon_r^2$.
Estimating $\epsilon_r\approx4$ for the effect of a substrate and the screening by the other carbon bands,
this is around $B_\text{c}\approx1.43$ T.

As $\hat V^\text{F}$ does not couple different spin and valley components, the trial state and its energy [Eq.~(\ref{bilayeronegeneral})] generalize trivially with a
component-specific parameter set $z_{0\xi\sigma}$, $z_{1\xi\sigma}$.
The states with several partially filled components are disfavored at any integral filling $\overline\nu$ \cite{supp}.
Thus at $\nu=-2,0,2$ there are only completely filled and empty components, confirming Hund's rule.\cite{Hund}
At $\nu=\pm3,\pm1$, one partially filled component has the structure given previously for the spinless case [Fig.~\ref{params1}(a)]; $(\nu+3)/2$ components are filled, and the rest are empty.

For the bifurcation it was essential that a proper QHF structure in the ZLB removed the inequivalence of the
sublattices that correspond to the spinor components.
As this needed half-filling of an orbital sector, we may expect similar phenomena at $\overline\nu=\nu+8=2$
filling of four-layer graphene in ABCA stacking.
Again, we assume the spin and valley degrees of freedom have ordered,
and seek the ground state as
$\Psi^{(4)}_{2}
=\prod_q\left(\sum_{i<j}w_{ij}\hat c^\dag_{iq}\hat c^\dag_{jq}\right)\Psi^{(4)}_0$,
with skew-symmetric $w_{ij}$, and $\sum_{i<j}|w_{ij}|=1$.
The ground-state energy is
\begin{multline}
\label{abcatwo}
\frac{E^{(4)}_{2}}{N_\phi}=\frac{E^{(4)}_0}{N_\phi}
-\frac{1}{2}\sum_{i=0}^3X_{ii}C_{ii}^2
-\sum_{i<j}\left(X_{ij}C_{ii}C_{jj}+X_{iijj}|C_{ij}|^2\right)-\\
-2X_{0121}\Re(C_{21}C_{01})-2X_{0213}\Re(C_{01}C_{32}+C_{31}C_{02})-\\
-2X_{1223}\Re(C_{32}C_{12})-\sum_{m=4}^\infty\sum_{i=0}^3X_{im}C_{ii}w^+(\theta_m,\phi_m),
\end{multline}
where $C_{ij} = \langle\hat c^\dag_{iq} \hat c_{jq}\rangle$ is detailed in Ref.~\onlinecite{supp}.
Let $w_{ij}=|w_{ij}|e^{i\phi_{ij}}$ with $\phi_{ij}$ real.
$E^{(4)}_{2}$ is unchanged by $\phi_{01}\to\phi_{01}+2\delta$, $\phi_{02}\to\phi_{02}+\delta$,
$\phi_{13}\to\phi_{13}-\delta$, and $\phi_{23}\to\phi_{23}-2\delta$.
The ground state manifold has U(1) symmetry.
Including spin and valley, this state occurs in one component at $\nu=\pm6,\pm2$;
other components are either full or empty.

The class of states with no preferred sublattice for the $|m|\ge4$ LL's will be called ``balanced''.
These fulfill the property $C_{ii}=\frac{1}{2}$, which is equivalent to
$|w_{01}|=|w_{23}|$, $|w_{02}|=|w_{13}|$, and $|w_{03}|=|w_{12}|$.
Then $\theta_m=0$, and we have six independent parameters to optimize.

In the weak-coupling limit $\beta^{(4)}\to0$ the ground state is balanced with
$w_{01}=w_{23}=0.311$, $w_{02}=0.504 e^{0.035i}$, $w_{03}=0.387 e^{0.012i}$, $w_{12}=0.387 e^{-0.014i}$ and $w_{13}=0.504 e^{0.04i}$.
With finite $\beta^{(4)}\gtrsim2.6$, the variational state restricted to the ``balanced'' subspace yields higher energies than the complete search [Fig.~\ref{params1}(b)],
which indicates a bifurcation as minima must come in particle-hole conjugate pairs $\{w_{ij},\theta_m,\phi_m\}$ and
$\{w_{01}\leftrightarrow w_{23},w_{02}\leftrightarrow-w_{13},w_{03}\leftrightarrow w_{12},\theta_m,\phi_m+\pi\}$.
$\beta^{(4)}_\text{c}\approx2.6$ corresponds to $B_\text{c}=86\text{ T}/\epsilon_r^{2/3}$;
using $\epsilon_r\approx4$, $B_\text{c}\approx34$ T.
As the two-band models overestimate the LL energies,\cite{Cote}
the mixing of the orbitals $\pm m$ may be less costly, reducing $\beta^{(4)}_\text{c}$,
possibly taking $B_\text{c}$ beyond the experimental range.

If a QHF state does not fill the orbital sector of the ZLB to one-half, balanced states do not exist.
Particle-hole symmetry is then manifest in the relation connecting the QHFs at $\nu$ and $-\nu$.
In particular, the mean-field ground state of ABC trilayer graphene at $\overline\nu=\nu+6=1,2$
and of the ABCA four-layer at $\overline\nu=\nu+8=1,3$ can be sought for in the form
$\Psi^{(s)}_{1}=\prod_q\left(\sum_{i=0}^{s-1}z_i\hat c^\dag_{iq}\right)\Psi^{(s)}_0$ and
$\Psi^{(s)}_{s-1}=\prod_q\left(\sum_{i=0}^{s-1}z_i\hat c_{iq}\right)
\left(\prod_{i=0}^{s-1}\hat c^\dag_{iq}\right)\Psi^{(s)}_0$, with $\sum_i|z_i|^2=1$.

\begin{figure}[htbp]
\begin{center}
\includegraphics[width=\columnwidth,keepaspectratio]{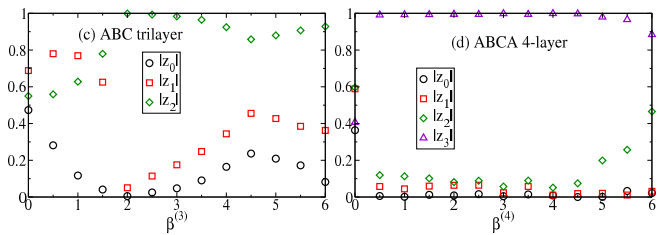}
\end{center}
\caption{\label{params2}(Color online)
The orbital amplitude parameters $|z_i|$ of the QHF state $\overline\nu=1,s-1$
in (a) ABC trilayer graphene and (b) ABCA four-layer graphene.
}
\end{figure}

By elementary algebra,\cite{supp} the phases of the $z_i$ parameters appear in the ground state energies
in a single term $-2X_{0121}\Re(z_0^\ast z_1^2z_2^\ast)$ for the trilayer and
$-2X_{0121}\Re(z_0^\ast z_1^2 z_2^\ast)-4X_{0213}\Re(z_0^\ast z_1z_2z_3^\ast)-2X_{1223}\Re(z_1^\ast z_2^2 z_3^\ast)$
for the four-layer, respectively.
The optimization of the phases imposes one and two constraints, respectively,
Thus the ground-state manifold has U(1) symmetry in each case.
Moreover, if $\{z_i,\theta_m,\phi_m\}$ optimizes the ground-state energy at $\overline\nu=1$, $\{z_i,\theta_m,\phi_m+\pi\}$ does the same at $\overline\nu=s-1$; the two states are related by intracomponent particle-hole symmetry.
For the magnitudes of the parameters, see Fig.~\ref{params2}.

With spin and valley, the ground state at $\nu=-s,0,s$ follows Hund's rule,\cite{Hund,phsymmbreaking}
at $\nu=-2s+1,-s+1,1$, and $s+1$ it has a partially filled component with the structure of $\Psi^{(s)}_1$,
and at $\nu=-s-1,-1,s-1$ and $2s-1$ it has a partially filled component with the structure of $\Psi^{(s)}_{s-1}$.
Particle-hole symmetry always holds.

\textit{Conclusion.}
When the filling factor is in the zero-energy Landau band, but not a multiple of the number of layers,
the QHF ground state is orbitally coherent with an orbital U(1) symmetry.
The orbital order should disappear in a Berezinsky-Kosterlitz-Thouless transition at some finite temperature.
Below that an orbital Goldstone mode is expected.
The bifurcation of the ground-state manifold at half-filling of an orbital sector must be generic in
rhombohedral graphene with an even number of layers; we have presented evidence for this in two- and four-layers.
Multilayers in standard Bernal stacking \cite{Bernal} can be mapped to a collection of monolayers and bilayers; the latter with a scaled parameter
$\gamma_1=2\cos\left(\pi/2 - m\pi/(2(s-1))\right)$ with $m=1,3,\dots (s-1)$ for $s$ even and $m=0,2,\dots (s-1)$ for $s$ odd.\cite{aba}
Therefore, bifurcations in the bilayer-like part generates complicated ground-state structures in these systems.
The analysis of these transitions is beyond mean-field theory, and is delegated to a future publication.

\section*{Acknowledgement}
This research was funded by the Hungarian Scientific Research Funds (Grant No.\ K105149) and C. T. was supported by the Hungarian Academy Sciences.
I thank K. Shizuya for helpful discussion.

\clearpage
\onecolumngrid

\renewcommand{\thefigure}{S\arabic{figure}}
\renewcommand{\thetable}{S\arabic{table}}
\setcounter{equation}{0}
\setcounter{figure}{0}

\section*{Supplementary Online Material}

\section{The range of validity of the two-band models}

Our study is based on the two-band model of chiral (rhombohedral) multilayer graphene, $\hat H^{(s)}_\xi$
defined in Eq.~(1) of the paper.
Trigonal warping and next-nearest neighbor hoppings, named $\gamma_3$, $\gamma_4$ and $\gamma_2$
in the terminology of the Slonczewski-Weiss-McClure model, have been neglected.
We discuss the appropriateness of these approximations for the kind of calculations that is presented in the paper.

\subsection{Splitting of the (ideally) zero-energy Landau band}

Trigonal warping introduces some mixing among the orbitals of the two-band model.
This changes the energy of the $n=1$ orbitals in the zero-energy Landau band (ZLB) of the bilayer,
the $n=1$ and $n=2$ orbitals of the ABC trilayer, etc.
I discuss this effect based on Ref.~18.

(i) Inspecting Fig.~5(b,d) in Ref.~18, the energy splitting of the zero-energy Landau band is
roughly
\begin{equation}
\frac{|E_0-E_1|}{B}=\frac{0.01\text{ eV}}{26\text{ T}}
\end{equation}
for the bilayer and
\begin{equation}
\frac{\max\{|E_0-E_2|,|E_0-E_1|\}}{B}=\frac{0.01\text{ eV}}{30\text{ T}}
\end{equation}
for the chiral trilayer.
The four-band and six-band models, respectively, predict about 15\% less splitting at high magnetic fields ($\approx 30$ T);
for small fields the six-band model predicts about 50\% less than the two-band model for the trilayer.
In comparison to the Coulomb energy scale, which is relevant to the exchange energy considerations of the paper,
\begin{equation}
|E_0-E_1|\left/\frac{e^2}{4\pi\epsilon\ell}\right.\approx\frac{\epsilon_r\sqrt{B\text{ [T]}}}{145.6}
\end{equation}
for the bilayer and
\begin{equation}
\max\{|E_0-E_2|,|E_0-E_1|\}\left/\frac{e^2}{4\pi\epsilon\ell}\right.\approx\frac{\epsilon_r\sqrt{B\text{ [T]}}}{168}
\end{equation}
for the ABC trilayer.

In the case of the bilayer, consider $\epsilon_r=4$ and $B=1.43$ T (which corresponds to the bifurcation at this
dielectric constant, one prediction our study), $|E_0-E_1|\left/\frac{e^2}{4\pi\epsilon\ell}\right.\approx0.032$;
this is a small quantitative correction.

In the trilayer, consider $\epsilon_r=4$ and $B=10$ T, a typical experimental setting.
Then $\max\{|E_0-E_2|,|E_0-E_1|\}\left/\frac{e^2}{4\pi\epsilon\ell}\right.\approx0.06$,
which might modify the coefficients of the $n=0,1,2$ orbitals (see Fig.~3), but it does not affect the qualitative
point I make for the trilayer, namely that Hund's rule is violated and particle-hole symmetry always holds.

\subsection{Distortion of the low-index Landau orbitals}

Trigonal warping starts to influence the structure of the Landau orbitals when the energy of the lowest Landau level
outside of the zero-energy band, of index $n=\pm s$, approaches the energy of the saddle point that
trigonal warping introduces in the low-energy spectrum (c.f.\ Ref.\ 14).
The latter energy scale is also called the Lifsitz transition energy, as the constant energy contour
becomes multiply connected below this energy.
This issue is different from the effect of trigonal warping on the spectrum; for the latter
it is more appropriate the compare the small-index Landau level energies with the energy scale that appears
in front of the trigonal warping term in the two-band model, c.f.\ Sec.~III.A.\ of Ref.~18.

For bilayer graphene, $E^{(2)}_{LiTr}=\frac{mv_3^2}{2}$,
where $v_3\approx 10^5$ m/s, $m=\frac{\gamma_1}{2v^2}$, $\gamma_1\approx0.4$ eV, and $v\approx 10^6$ m/s.
From $\hbar\omega_c\sqrt2= E^{(2)}_{LiTr}$ we get
$B=\frac{\gamma_1^2v_3^2}{8v^2\hbar e\sqrt2}\approx0.215$ T.
This is a rather small magnetic field. The interaction strength parameter,
introduced in the paper and Eq.~(\ref{intstrength}) below, is
$\beta^{(2)}=\frac{e^2}{4\pi\epsilon\ell}\left/
\frac{2\hbar v^2eB}{\gamma_1}\right.
\approx\frac{17}{\epsilon_r\sqrt{B\text{ [T]}}}$,
which, for the above magnetic field and a generous estimation $\epsilon_r\approx4$,
corresponds to $\beta^{(2)}\approx 9.2$.
This is definitely above the range $\beta^{(2)}\approx 3.5$ where the bifurcation should occur.

In Ref.~17 we calculated the overlap of the Landau
orbitals in the presence of trigonal warping (and other hoppings) with those without trigonal warping
in the four-band model of bilayer graphene.
In Fig.~2 the overlap is shown as the parameters $\gamma_3$, $\gamma_4$, and $\Delta'$
are tuned from zero to their literary value.
We found that trigonal warping is definitely the most important perturbation.
The change of the $n=1,\pm2,\pm3$ Landau orbitals
is significant for $B=0.1$ T, but rather small for $B=1$ T;
the change of higher energy orbitals is exptected to be even smaller.
This numerical finding is consistent with the above argument that trigonal warping becomes
significant below or near $0.2$ T.
Thus at and around the critical interaction strength, which corresponds to $B=1.43$ T at $\epsilon_r=4$,
the distortion of small-energy orbitals outside the zero-energy Landau band is rather small.

For ABC trilayer graphene, using Koshino and McCann's Ref.~14,
$E^{(3)}_{LiTr}=\frac{\gamma_1}{4}\left(\frac{v_3}{v}\right)^2$.
Setting the energy of the first excited Landau level $\frac{(2\hbar v^2eB)^{3/2}}{\gamma_1^2}\sqrt6$ equal to the
Lifsitz transition energy $E^{(3)}_{LiTr}$ we get
$B=\left(\frac{v_3}{v}\right)^{4/3}
\frac{\gamma_1^2}{\hbar v^2e 2^{8/3}3^{1/3}}\approx1.232$ T.
This is no longer small, but typical experiments are performed in somewhat higher magnetic fields.
The corresponding interaction strength parameter is
$\beta^{(3)}=\frac{e^2}{4\pi\epsilon\ell}\left/\frac{(2\hbar v^2eB)^{3/2}}{\gamma_1^2}\right.
\approx\frac{188}{\epsilon_r B\text{ [T]}}$,
which, for the above magnetic field and $\epsilon_r\approx4$,
corresponds to $\beta^{(3)}\approx 38$.
Thus our figures, plotted up to $\beta^{(3)}=6$, are on the safe side.

For the ABCA four-layer, using Eq.~(14) of Ref.~14, using $\gamma_2=0$,
the saddle point separating the Dirac point at the origin from one of the three
side Dirac points can be found to be at $E^{(4)}_{LiTr}=4\gamma_1\left(\frac{v_3}{v}\right)^4$.
This is a tiny energy, but the Landau levels are also rather dense.
Setting the energy of the first excited Landau level $\frac{(2\hbar v^2eB)^2}{\gamma_1^3}\sqrt{24}$ equal to the
Lifsitz transition energy $E^{(4)}_{LiTr}$, we get
$B=(24)^{1/4}\left(\frac{v_3}{v}\gamma_1\right)^2\frac{1}{\hbar v^2e}\approx5.38$ T.
The corresponding interaction strength is
$\beta^{(4)}=\frac{e^2}{4\pi\epsilon\ell}\left/\frac{(2\hbar v^2eB)^{2}}{\gamma_1^3}\right.
\approx\frac{2073}{\epsilon_r B^{3/2}\text{ [T]}}$,
which, for the above magnetic field and $\epsilon_r\approx4$,
corresponds to $\beta^{(3)}\approx 41.5$.

\subsection{Single-particle energies beyond the zero-energy Landau band}

The two-band model typically overestimates the single-particle energies outside of the ZLB, c.f.\ Ref.~18.
This in turn overestimates the cost of mixing the index $-m$ and $+m$ orbitals to minimize the exchange energy.
In the cases where bifurcation occurs in the ground state manifold, this leads to an
overestimation of the critical interaction strength $\beta^{(2)}$ or $\beta^{(4)}$.
In a fixed dielectric environment this leads to an underestimation of the critical magnetic field.
As estimated in the paper, the critical fields are $B_\text{c}\approx1.43$ T for the bilayer at
$\nu=-3$ and $B_\text{c}\approx34$ T for the chiral four-layer at $\nu=-7$.
A moderate upward shift of the first is harmless, but in the second case it may render the
bifurcation a theoretical possibility.

\subsection{Separation from the split bands}

We estimate the cutoff Landau level $M$ under which the two-band model is valid from
setting its energy equal to the interlayer nearest neighbor hopping $\gamma_1$.
Recall that two split bands start at $\pm\gamma_1$.
Thus, for the bilayer,
\begin{gather*}
\hbar\omega_c\sqrt{M(M-1)}=\frac{2\hbar v^2eB}{\gamma_1}\sqrt{M(M-1)}=\gamma_1,\\
M(M-1)=\left(\frac{\gamma_1^2}{2\hbar v^2eB}\right)^2 \approx \left(\frac{122}{B\text{ [T]}}\right)^2.
\end{gather*}
Hence $M\lesssim\frac{122}{B\text{ [T]}}+\frac{1}{2}$.
For the ABC trilayer,
\begin{gather*}
\frac{(2\hbar v^2eB)^{3/2}}{\gamma_1^2}\sqrt{M(M-1)(M-2)}=\gamma_1,\\
M(M-1)(M-2)=\left(\frac{\gamma_1^2}{2\hbar v^2eB}\right)^3 \approx \left(\frac{122}{B\text{ [T]}}\right)^3.
\end{gather*}
Hence $M\lesssim\frac{122}{B\text{ [T]}}+1$.
For the ABCA four-layer,
\begin{gather*}
\frac{(2\hbar v^2eB)^{2}}{\gamma_1^3}\sqrt{M(M-1)(M-2)(M-3)}=\gamma_1,\\
M(M-1)(M-2)(M-3)=\left(\frac{\gamma_1^2}{2\hbar v^2eB}\right)^4 \approx \left(\frac{122}{B\text{ [T]}}\right)^4.
\end{gather*}
Hence $M\lesssim\frac{122}{B\text{ [T]}}+\frac{3}{2}$.

Thus, the Landau level cutoff can be safely estimated for all cases we are interested in as
\begin{equation}
M\lesssim\frac{120}{B\text{ [T]}}.
\end{equation}

\section{The mean-field Hamiltonian}

Using the Landau gauge $\mathbf A=\mathbf{\hat y}Bx$, the Landau orbitals of the \textit{conventional two-dimensional electron gas} are
\begin{equation}
\label{etaeq}
\eta_{nq}(\mathbf r)=\frac{e^{iqx-\left(y/\ell-q\ell\right)^2/2}}{\sqrt{2\pi\sqrt\pi 2^n n!\ell}}H_n\left( \frac{y}{\ell}-q\ell \right),
\end{equation}
where $H_n$ is a Hermite-polynomial.
I often use the matrix elements of the plane wave,
\begin{equation}
\langle n',q'|e^{-i\mathbf k\cdot\mathbf r}|nq\rangle=\delta(q-q'-k_y)F_{n'n}(\mathbf k)e^{-ik_x\ell^2(q+q')/2},
\end{equation}
where I define
\begin{equation}
\label{twodegf}
F_{n'n}(\mathbf k)=\sqrt\frac{n!}{(n')!}\left(\ell\frac{k_y-ik_x}{\sqrt2}\right)^{n'-n}L_n^{n'-n}\left(\frac{k^2\ell^2}{2}\right)e^{-k^2\ell^2/4},
\end{equation}
if $n'\ge n$, else $F_{nn'}(\mathbf q)=F^\ast_{n'n}(-\mathbf q)$. $L^m_n(z)$ is an associated Laguerre polynomial.
Then the density operator is
\begin{equation}
\label{twodegdensity}
\hat\rho_{\sigma\sigma'}(\mathbf k)=\int d^2r e^{-i\mathbf k\cdot\mathbf r}\Psi^\dag_\sigma(\mathbf r)\Psi_{\sigma'}(\mathbf r)
=\sum_{n,n'}\int dp F_{n'n}(\mathbf k)e^{-ik_xp\ell^2}\hat c^\dag_{n',p-k_y/2,\sigma}\hat c_{n,p+k_y/2,\sigma'}.
\end{equation}
The interaction part of the Hamiltonian is
\begin{multline}
\hat V=\frac{e^2}{8\pi\epsilon}\sum_{\sigma,\sigma'}
\int\frac{d^2k}{(2\pi)^2}:\hat\rho_{\sigma\sigma}(\mathbf k)\hat\rho_{\sigma'\sigma'}(-\mathbf k):\frac{2\pi}{k}=\\
=\frac{e^2}{8\pi\epsilon}\sum_{n_1n_2n_3n_4}\int dp_1\int dp_2\sum_{\sigma,\sigma'}
\int\frac{d^2k}{(2\pi)^2}F_{n_1n_4}(\mathbf k)F_{n_2n_3}(-\mathbf k)\frac{2\pi}{k}\times\\
\times\hat c^\dag_{n_1p_1\sigma}\hat c^\dag_{n_2p_2\sigma'}\hat c_{n_3,p_2-k_y,\sigma'}\hat c_{n_4,p_1+k_y,\sigma}e^{ik_x\ell^2(p_2-p_1-k_y)}.
\end{multline}
Now, if the basis states have some spinorial structure,
\begin{equation}
\Psi_{nq}(\mathbf r)=\begin{pmatrix}
A_n\eta_{nq}(\mathbf r)\\
B_n\eta_{n-sq}(\mathbf r)
\end{pmatrix},
\end{equation}
$\hat c^\dag_{nq}\hat c^\dag_{n'q'}$ creates a two-particle state with
$\text{Det}\begin{pmatrix}
A_n\eta_{nq}(\mathbf r_1) & A_{n'}\eta_{n'q'}(\mathbf r_1)\\
A_n\eta_{nq}(\mathbf r_2) & A_{n'}\eta_{n'q'}(\mathbf r_2)
\end{pmatrix}$ in one sublattice, and
$\text{Det}\begin{pmatrix}
B_n\eta_{n-s,q}(\mathbf r_1) & B_{n'}\eta_{n'-s,q'}(\mathbf r_1)\\
B_n\eta_{n-s,q}(\mathbf r_2) & B_{n'}\eta_{n'-s,q'}(\mathbf r_2)
\end{pmatrix}$ in the other sublattice.
Thus the interaction becomes
\begin{multline}
\hat V=\frac{e^2}{8\pi\epsilon}\sum_{n_1n_2n_3n_4}\int dp_1\int dp_2\sum_{\sigma,\sigma'}
\int\frac{d^2k}{(2\pi)^2}\\
\left(
A^\ast_{n_1}A^\ast_{n_2}A_{n_3}A_{n_4} F_{n_1n_4}(\mathbf k)F_{n_2n_3}(-\mathbf k)V_{11}(\mathbf k)
+A^\ast_{n_1}B^\ast_{n_2}B_{n_3}A_{n_4} F_{n_1n_4}(\mathbf k)F_{n_2-s,n_3-s}(-\mathbf k)V_{12}(\mathbf k)+\right.\\
+B^\ast_{n_1}A^\ast_{n_2}A_{n_3}B_{n_4} F_{n_1-s,n_4-s}(\mathbf k)F_{n_2n_3}(-\mathbf k)V_{21}(\mathbf k)+
\left.+B^\ast_{n_1}B^\ast_{n_2}B_{n_3}B_{n_4} F_{n_1-s,n_4-s}(\mathbf k)F_{n_2-s,n_3-s}(-\mathbf k)V_{22}(\mathbf k)\right)\\
\hat c^\dag_{n_1p_1\sigma}\hat c^\dag_{n_2p_2\sigma'}\hat c_{n_3,p_2-k_y,\sigma'}\hat c_{n_4,p_1+k_y,\sigma}e^{ik_x\ell^2(p_2-p_1-k_y)},
\end{multline}
where, using the distance $d=(s-1)0.335$ nm between the layers,
\begin{equation}
V_{11}(\mathbf k)=V_{22}(\mathbf k)=\frac{2\pi}{k},\quad\quad
V_{12}(\mathbf k)=V_{21}(\mathbf k)=\frac{2\pi}{k}e^{-d|q|}.
\end{equation}
As $d\ll\ell$, the difference between $V_{11}=V_{22}$ and $V_{12}=V_{21}$ can be neglected, and
\begin{multline}
\hat V=\frac{e^2}{8\pi\epsilon}\sum_{n_1n_2n_3n_4}\int dp_1\int dp_2\sum_{\sigma,\sigma'}
\int\frac{d^2k}{(2\pi)^2}\widetilde F_{n_1n_4}(\mathbf k)\widetilde F_{n_2n_3}(-\mathbf k)\frac{2\pi}{k}\\
\hat c^\dag_{n_1p_1\sigma}\hat c^\dag_{n_2p_2\sigma'}\hat c_{n_3,p_2-k_y,\sigma'}\hat c_{n_4,p_1+k_y,\sigma}e^{ik_x\ell^2(p_2-p_1-k_y)},
\end{multline}
with\cite{foot1}
\begin{equation}
\widetilde F_{n'n}(\mathbf k)= A_nA_{n'}F_{n'n}(\mathbf k) + B_nB_{n'}F_{n'-s,n-s}(\mathbf k).
\end{equation}
With the particular choice of orbitals of the minimal model of rhombohedral $s$-layer graphene, we get $\widetilde F_{nn'}(\mathbf k)=F^{(s)}_{nn'}(\mathbf k)$,
\begin{equation}
F^{(s)}_{nn'}(\mathbf k)=\left\{
\begin{array}{cl}
F_{nn'}(\mathbf k) & \text{if $0\le n<s$ and $0\le n'<s$},\\
\frac{1}{\sqrt 2}F_{|n||n'|}(\mathbf k) & \text{if either $0\le n<s$ or $0\le n'<s$},\\
\frac{1}{2}\left(F_{|n||n'|}(\mathbf k)+\text{sgn}(nn')F_{|n|-s,|n'|-s}(\mathbf k)\right) & \text{if $|n|\ge s$ and $|n'|\ge s$},\\
\text{undefined} & \text{if $-s<n<0$ or $-s<n'<0$.}
\end{array}
\right.
\end{equation}

By standard mean-field decomposition $\hat V^{\text{HF}} = \hat V^{\text{H}} + \hat V^{\text{F}}$, with
\begin{multline}
\frac{\hat V^\text{H}}{N_\phi}=-\frac{e^2}{8\pi\epsilon}
\sum_{nn'mm'}\sum_{\xi\xi'\sigma\sigma'}\int dp
\int\frac{d^2k}{(2\pi)^2}F^{(s)}_{nm}(\mathbf k)F^{(s)}_{m'n'}(-\mathbf k)\frac{2\pi}{k}\times\\
\times\left[\left\langle \hat c^\dag_{m'p\xi'\sigma'}
\hat c_{n'p\xi'\sigma'}
\right\rangle 
\hat c^\dag_{nq\xi\sigma}
\hat c_{mq\xi\sigma}
+
\left\langle
\hat c^\dag_{np\xi\sigma}
\hat c_{mp\xi\sigma}
\right\rangle
\hat c^\dag_{m'q\xi'\sigma'}
\hat c_{n'q\xi'\sigma'}
-
\left\langle
\hat c^\dag_{np\xi\sigma}
\hat c_{mp\xi\sigma}
\right\rangle
\left\langle
\hat c^\dag_{m'q\xi'\sigma'}
\hat c_{n'q\xi'\sigma'}
\right\rangle
\right],
\end{multline}
and
\begin{multline}
\label{meanfield2}
\frac{\hat V^\text{F}}{N_\phi}=-\frac{e^2}{4\pi\epsilon}\sum_{nn'mm'}\sum_{\xi\xi'\sigma\sigma'}
\int\frac{d^2k}{(2\pi)^2}F^{(s)}_{nm}(\mathbf k)F^{(s)}_{m'n'}(-\mathbf k)\frac{2\pi}{k}\times\\
\times\left[\left\langle \hat c^\dag_{m'p\xi\sigma}\hat c_{mp\xi'\sigma'}\right\rangle \hat c^\dag_{n,p-k_y,\xi'\sigma'}\hat c_{n'_,p-k_y,\xi\sigma}-
\frac{1}{2}\left\langle \hat c^\dag_{m'p\xi\sigma}\hat c_{mp\xi'\sigma'}\right\rangle
\left\langle\hat c^\dag_{n,p-k_y,\xi'\sigma'}\hat c_{n'_,p-k_y,\xi\sigma}\right\rangle
\right],
\end{multline}
where $p$ is arbitrary but fixed, $N_\phi$ is the number of flux quanta piercing the sample.
Notice that the angular integral in Eq.~(\ref{meanfield2}) vanishes unless $|n|-|m|=|n'|-|m'|$.

For future reference let us introduce the following exchange energy constants:
\begin{align}
X_{n_1n_2n_3n_4}&=\frac{e^2}{4\pi\epsilon}\int\frac{d^2k}{(2\pi)^2}
\frac{2\pi}{k}F_{n_1n_2}(\mathbf k)F^\ast_{n_3n_4}(\mathbf k),\label{exint}\\
X_{n'n}=X_{n'nn'n}&=\frac{e^2}{4\pi\epsilon}
\int\frac{d^2k}{(2\pi)^2}\frac{2\pi}{k}|F_{n'n}(\mathbf k)|,\label{exchange}\\
X^{(s)}_{n'n}=X_{n'nn'-s,n-s}&=\frac{e^2}{4\pi\epsilon}
\int\frac{d^2k}{(2\pi)^2}\frac{2\pi}{k}F_{n'n}(\mathbf k)F^\ast_{n'-s,n-s}(\mathbf k).
\end{align}
Closed form expressions for $X_{n'n}$ and $X^{(s)}_{n'n}$ are provided in Section \ref{secexchange},
and relevant special cases are given in Table \ref{xtable}.
Notice that all of these expressions scale with $\frac{e^2}{4\pi\epsilon\ell}$, which is the natural scale of the
interaction energy in these systems.

Recall that the single-particle (Landau level) spectrum is
\begin{equation}
\epsilon^{(s)}_n=\text{sgn}(n)\varepsilon_s\sqrt{\prod_{i=0}^{s-1}(|n|-i)},
\end{equation}
with
\begin{equation}
\varepsilon_s\equiv\frac{(2\hbar v^2eB)^{s/2}}{\gamma_1^{(s-1)}}.
\end{equation}
Let us characterize the relative strength of the interaction by
\begin{equation}
\label{intstrength}
\beta^{(s)}=\frac{e^2}{4\pi\epsilon\ell}\Bigg/\varepsilon_s.
\end{equation}

\subsection{Simplifications for uniform liquid states}

The mean-field theory is greatly simplified for uniform liquid states, i.e.,
if we assume translational and rotational invariance of the density.
First I discuss the simpler case of the conventional two-dimensional electron gas.
The expectation value of the density [Eq.~(\ref{twodegdensity})] can be written as
\begin{equation}
\label{rhogdc}
\langle\hat\rho_{\sigma\sigma'}(\mathbf q)\rangle=\sum_{n,n'=0}^{\infty}F_{n'n}(\mathbf q)\Delta^{nn'}_{\sigma\sigma'}(\mathbf q),
\end{equation}
where the function $F_{n'n}(\mathbf q)$ was defined in Eq.~(\ref{twodegf}), and the \textit{guiding center density} is defined as follows:
\begin{equation}
\Delta^{nn'}_{\sigma\sigma'}(\mathbf q)=\int dp e^{-iq_xp\ell^2}\langle\hat c^\dag_{n',p-q_y/2,\sigma'}\hat c_{n,p+q_y/2,\sigma}\rangle.
\end{equation}
Assuming translational invariance, the expectation value $\langle\hat c^\dag_{n',p-q_y/2,\sigma'}\hat c_{n,p+q_y/2,\sigma}\rangle$
cannot depend on $p$; hence
\begin{equation}
\label{deltax}
\Delta^{nn'}_{\sigma\sigma'}(\mathbf q)\propto\delta(q_x).
\end{equation}
Rotating the coordinate system by $\pi/2$ in either direction and assuming the density does not change,
\begin{equation}
\label{deltay}
\Delta^{nn'}_{\sigma\sigma'}(\mathbf q)\propto\delta(q_y).
\end{equation}

Now, let $\langle\hat\rho_{\sigma\sigma'}(\mathbf q)\rangle=\overline\rho\delta(\mathbf q)$ and
$\Delta^{nn'}_{\sigma\sigma'}(\mathbf q)=\overline\Delta^{nn'}_{\sigma\sigma'}\delta(\mathbf q)$.
Let us consider an ensemble of rotationally invariant systems that approach a homogeneous (translationally
invariant) system in the thermodynamic limit.
Consider a state with the density
\begin{equation}
\langle\hat\rho_{\sigma\sigma'}(\mathbf q)\rangle=\overline\rho\frac{e^{-\frac{q^2}{4\epsilon}}}{4\pi\epsilon}.
\end{equation}
In the $\epsilon\to+0$ limit this density converges to $\overline\rho\delta(\mathbf q)$, i.e., the
density of a homogeneous system.\cite{foot2}
For finite but small $\epsilon$ the density is a sharp peak in momentum space, or a large but finite droplet in
real space. Similarly,
\begin{equation}
\Delta^{nn'}_{\sigma\sigma'}(\mathbf q)=\overline\Delta^{nn'}_{\sigma\sigma'}\frac{e^{-\frac{q^2}{4\epsilon}}}{4\pi\epsilon}.
\end{equation}
This procedure is analogous to the treatment of the Laughlin state for the fractional quantum Hall effect at
filling factor $\nu=\frac{1}{2m+1}$, where the variational state is first studied for a finite number of
particles and a corresponding fixed total angular momentum.
This is a liquid state confined to a finite disk; the homogeneous liquid state is interpreted as the thermodynamic limit of such droplets.
We follow this procedure to exploit rotational invarience; if we considered the homogeneous system
directly, the density would be proportional to $\delta(\mathbf q)$ and all propositions capturing rotational
invariance in momentum space would be vacuous.
Consider the circle $\mathcal C=\langle Q\cos\theta,Q\sin\theta\rangle$ at fixed $Q$ and variable $0\le\theta\le 2\pi$.
Multiply Eq.~(\ref{rhogdc}) by $e^{i\theta k}$ and integrate over $\mathcal C$:
\begin{gather}
\overline\rho\frac{e^{-\frac{Q^2}{4\epsilon}}}{2\epsilon}\delta_{k,0}=
\sum_{n,n'=0}^{\infty}F_{n'n}(\mathbf Q)\overline\Delta^{nn'}_{\sigma\sigma'}
\frac{e^{-\frac{Q^2}{4\epsilon}}}{2\epsilon}\delta_{n-n',k}
=\sum_{n'=0}^{\infty}F_{n'+k,n'}(\mathbf Q)\overline\Delta^{n'+k,n'}_{\sigma\sigma'}
\frac{e^{-\frac{Q^2}{4\epsilon}}}{2\epsilon},\nonumber\\
\overline\rho\delta_{k,0}=\sum_{n'=0}^{\infty}F_{n'+k,n'}(\mathbf Q)\overline\Delta^{n'+k,n'}_{\sigma\sigma'}.\label{restrict}
\end{gather}
This holds for all $Q$ only if the coefficient of all $F_{n'+k,n'}(\mathbf Q)$ vanishes for $k\neq0$, i.e.\
$\overline\Delta^{n'+k,n'}_{\sigma\sigma'}\propto\delta_{k,0}$ or
$\overline\Delta^{nn'}_{\sigma\sigma'}\propto\delta_{n,n'}$.

Notice that in the $\epsilon\to+0$ limit the right hand side of Eq.~(\ref{restrict}) goes over to
$\delta(\mathbf Q)F_{n'+k,n'}(\mathbf Q)=\delta(\mathbf Q)F_{n'+k,n'}(0)=\delta(\mathbf Q)\delta_{k,0}$.
That is the reason why a non-zero constant value of the diagonal coefficient $\overline\Delta^{n'n'}_{\sigma\sigma'}$
becomes possible.

We still have to show that the exchange interaction can be written in terms of the guiding center density.
In this part of the argument it is more convenient to consider a finite $L_x\times L_y$ sample with periodic boundary
conditions; taking the thermodynamic limit at the end is straightforward.
Consider the exchange part of the mean-field Hamiltonian,
\begin{multline}
\hat V^\text{F}=-\frac{e^2}{4\pi\epsilon}\sum_{n_1n_2n_3n_4}\sum_{\sigma\sigma'}\sum_{p_1p_2}
\frac{1}{L_xL_y}\sum_{\mathbf k}
F_{n_1n_4}(\mathbf k)F_{n_2n_3}(-\mathbf k)\frac{2\pi}{k}e^{ik_x\ell^2(p_2-p_1)}\\
\left[\left\langle \hat c^\dag_{n_1,p_1-k_y/2,\sigma}\hat c_{n_3,p_2-k_y/2,\sigma'}\right\rangle
\hat c^\dag_{n_2,p_2+k_y/2,\sigma'}\hat c_{n_4,p_1+k_y/2,\sigma}
-\frac{1}{2}\left\langle \hat c^\dag_{n_1,p_1-ky/2,\sigma}\hat c_{n3,p_2-k_y/2,\sigma'}\right\rangle
\left\langle\hat c^\dag_{n_2,p_2+k_y/2,\sigma'}\hat c_{n_4,p_1+k_y/2,\sigma}\right\rangle
\right].\label{gdcfrom}
\end{multline}
Introduce $P=\frac{p_1+p_2-k_y}{2}$ and $\Delta p=p_1-p_2$:
\begin{multline}
\hat V^\text{F}=-\frac{e^2}{4\pi\epsilon}\sum_{n_1n_2n_3n_4}\sum_{\sigma\sigma'}\sum_{P,\Delta p}
\frac{1}{L_xL_y}\sum_{\mathbf k}
F_{n_1n_4}(\mathbf k)F_{n_2n_3}(-\mathbf k)\frac{2\pi}{k}e^{-ik_x\ell^2\Delta p}\\
\left[\left\langle \hat c^\dag_{n_1,P+\Delta p/2,\sigma}\hat c_{n_3,P-\Delta p/2,\sigma'}\right\rangle
\hat c^\dag_{n_2,P-\Delta p/2+k_y,\sigma'}\hat c_{n_4,P+\Delta p/2+k_y,\sigma}-\right.\\
\left.-\frac{1}{2}\left\langle \hat c^\dag_{n_1,P+\Delta p/2,\sigma}\hat c_{n3,P-\Delta p/2,\sigma'}\right\rangle
\left\langle\hat c^\dag_{n_2,P-\Delta p/2+k_y,\sigma'}\hat c_{n_4,P+\Delta p/2+k_y,\sigma}\right\rangle
\right].
\end{multline}
Substitute $k_y\to t=P+k_y$,
\begin{multline}
\hat V^\text{F}=-\frac{e^2}{4\pi\epsilon L_xL_y}\sum_{n_1n_2n_3n_4}\sum_{\sigma\sigma'}\sum_{P,\Delta p}
\sum_{t,k_x}
F_{n_1n_4}(k_x,t-P)F_{n_2n_3}(-k_x,P-t)\frac{2\pi}{\sqrt{k_x^2+(P-t)^2}}e^{-ik_x\ell^2\Delta p}\\
\left[\left\langle \hat c^\dag_{n_1,P+\Delta p/2,\sigma}\hat c_{n_3,P-\Delta p/2,\sigma'}\right\rangle
\hat c^\dag_{n_2,t-\Delta p/2,\sigma'}\hat c_{n_4,t+\Delta p/2,\sigma}-
\frac{1}{2}\left\langle \hat c^\dag_{n_1,P+\Delta p/2,\sigma}\hat c_{n3,P-\Delta p/2,\sigma'}\right\rangle
\left\langle\hat c^\dag_{n_2,t-\Delta p/2,\sigma'}\hat c_{n_4,t+\Delta p/2,\sigma}\right\rangle
\right].
\end{multline}
Let us introduce the Fourier transform of a subformula by
\begin{equation}
F_{n_1n_4}(\mathbf Q)F_{n_2n_3}(-\mathbf Q)\frac{2\pi}{|\mathbf Q|}=\int dx\int dy e^{-ixQ_x-iyQ_y}G^{n_1n_2n_3n_4}(x,y).
\end{equation}
Then
\begin{multline}
\hat V^\text{F}=-\frac{e^2}{4\pi\epsilon L_xL_y}\int dx\int dy
\sum_{n_1n_2n_3n_4}\sum_{\sigma\sigma'}\sum_{P,\Delta p,t,k_x}
G^{n_1n_2n_3n_4}(x,y)e^{-i(x+\ell^2\Delta p)k_x-iy(t-P)}\\
\left[\left\langle \hat c^\dag_{n_1,P+\Delta p/2,\sigma}\hat c_{n_3,P-\Delta p/2,\sigma'}\right\rangle
\hat c^\dag_{n_2,t-\Delta p/2,\sigma'}\hat c_{n_4,t+\Delta p/2,\sigma}-
\frac{1}{2}\left\langle \hat c^\dag_{n_1,P+\Delta p/2,\sigma}\hat c_{n3,P-\Delta p/2,\sigma'}\right\rangle
\left\langle\hat c^\dag_{n_2,t-\Delta p/2,\sigma'}\hat c_{n_4,t+\Delta p/2,\sigma}\right\rangle
\right].
\end{multline}
Using the definition of the guiding center density,
\begin{multline}
\hat V^\text{F}=-\frac{e^2}{4\pi\epsilon L_xL_y}\int dx\int dy
\sum_{n_1n_2n_3n_4}\sum_{\sigma\sigma'}\sum_{\Delta p,t,k_x}
G^{n_1n_2n_3n_4}(x,y)e^{-i(x+\ell^2\Delta p)k_x+iyt}\\
\Delta^{n_3n_1}_{\sigma'\sigma}\left(-y/\ell^2,-\Delta p\right)
\left[\hat c^\dag_{n_2,t-\Delta p/2,\sigma'}\hat c_{n_4,t+\Delta p/2,\sigma}-
\frac{1}{2}\left\langle\hat c^\dag_{n_2,t-\Delta p/2,\sigma'}\hat c_{n_4,t+\Delta p/2,\sigma}\right\rangle
\right].
\end{multline}
Now, using $\Delta^{n_3n_1}_{\sigma'\sigma}\left(-y/\ell^2,-\Delta p\right)\propto\delta_{y,0}\delta_{\Delta p,0}$,
\begin{multline}
\hat V^\text{F}=-\frac{e^2}{4\pi\epsilon L_x}\int dx
\sum_{n_1n_2n_3n_4}\sum_{\sigma\sigma'}\sum_{t,k_x}
G^{n_1n_2n_3n_4}(x,0)e^{-ixk_x}
\Delta^{n_3n_1}_{\sigma'\sigma}(0)
\left[
\hat c^\dag_{n_2t\sigma'}\hat c_{n_4t\sigma}-
\frac{1}{2}\left\langle\hat c^\dag_{n_2t\sigma'}\hat c_{n_4t\sigma}\right\rangle
\right]=\\
=-\frac{e^2}{4\pi\epsilon}
\sum_{n_1n_2n_3n_4}\sum_{\sigma\sigma'}\sum_{t}
G^{n_1n_2n_3n_4}(0,0)
\Delta^{n_3n_1}_{\sigma'\sigma}(0)
\left[\hat c^\dag_{n_2t\sigma'}\hat c_{n_4t\sigma}-
\frac{1}{2}\left\langle\hat c^\dag_{n_2t\sigma'}\hat c_{n_4t\sigma}\right\rangle
\right].
\end{multline}
Now use that $\Delta^{n_3n_1}_{\sigma'\sigma}(0)\propto\delta_{n_1.n_3}$ and that
$G^{n_1n_2n_3n_4}(0,0)=X_{n_1n_4n_3n_2}$ vanishes unless $n_1-n_4=n_3-n_2$,
\begin{equation}
\hat V^\text{F}=-\frac{e^2}{4\pi\epsilon}
\sum_{mn}\sum_{\sigma\sigma'}\sum_{t}X_{mnnm}
\Delta^{mm}_{\sigma'\sigma}(0)
\left[\hat c^\dag_{nt\sigma'}\hat c_{nt\sigma}-
\frac{1}{2}\left\langle\hat c^\dag_{nt\sigma'}\hat c_{nt\sigma}\right\rangle
\right].\label{gdcto}
\end{equation}
That is, the mean-field interaction does not mix Landau levels for uniform liquid states in the conventional two-dimensional
electron gas.

We now generalize the above argument for the case of the minimal model of rhombohedral bilayer graphene, as defined in Eq.~(1) of the paper.
The density operator
\begin{equation}
\hat\rho_{\xi\sigma,\xi'\sigma'}(\mathbf r,\mathbf r')=\Psi(\mathbf r)\otimes\Psi^\dag(\mathbf r')
\end{equation}
is in general a $2\times2$ matrix in sublattice space.

If we expand the field operators in the basis of Landau orbitals $\Psi_{nq\xi\sigma}$
(defined in the paper),
the $0\le n,n'<s$ state combinations contribute to the density operator with matrix terms like
\[
\begin{pmatrix}
\eta_{nq}(\mathbf r)\eta^\ast_{n'q}(\mathbf r') & 0\\
0 & 0
\end{pmatrix};
\]
the $0\le n<s,|n'|\ge s$ state combinations contribute with matrix terms like
\[\frac{1}{\sqrt2}
\begin{pmatrix}
\eta_{nq}(\mathbf r)\eta^\ast_{|n'|q}(\mathbf r') & \eta_{nq}(\mathbf r)\eta^\ast_{|n'|-s,q}(\mathbf r')\text{sgn}(n')\\
0 & 0
\end{pmatrix};
\]
the $0\le n'<s,|n|\ge s$ state combinations do that with matrix terms like
\[\frac{1}{\sqrt2}
\begin{pmatrix}
\eta_{|n|q}(\mathbf r)\eta^\ast_{n'q}(\mathbf r') & 0 \\
\eta_{|n|-s,q}(\mathbf r)\eta^\ast_{n',q}(\mathbf r')\text{sgn}(n) & 0
\end{pmatrix};
\]
and, finally, the $|n'|,|n|\ge s$ state combinations contribute to $\hat\rho_{\xi\sigma,\xi'\sigma'}(\mathbf r,\mathbf r')$
with matrix terms like
\[\frac{1}{2}
\begin{pmatrix}
\eta_{|n|q}(\mathbf r)\eta^\ast_{|n'|q}(\mathbf r') & \eta_{|n|q}(\mathbf r)\eta^\ast_{|n'|-s,q}(\mathbf r')\text{sgn}(n')\\
\eta_{|n|-s,q}(\mathbf r)\eta^\ast_{|n'|q}(\mathbf r')\text{sgn}(n) &
\eta_{|n|-s,q}(\mathbf r)\eta^\ast_{|n'|-s,q}(\mathbf r')\text{sgn}(n'n)
\end{pmatrix}.
\]
Taking the Fourier transform, we obtain 
\begin{equation}
\hat\rho_{\xi\sigma,\xi\sigma'}(\mathbf q)=\sum'_{n,n'}\mathbf F_{n'n}(\mathbf q)
\int dp e^{-iq_xp\ell^2}\hat c^\dag_{n',p-q_y/2,\sigma'}\hat c_{n,p+q_y/2,\sigma},
\end{equation}
where the summation is restricted to exclude the nonexisting Landau level indices $-s+1,\dots,-1$, and
\begin{equation}
\mathbf F_{n'n}(\mathbf q)=\left\{
\begin{array}{ll}
\begin{pmatrix}
F_{n'n}(\mathbf q) & 0 \\
0 & 0
\end{pmatrix},
&\text{if $0\le n,n'<s$;}\\
\frac{1}{\sqrt2}
\begin{pmatrix}
F_{n'|n|}(\mathbf q) & \text{sgn}(n)F_{n'|n|-s}(\mathbf q) \\
0 & 0
\end{pmatrix},
&\text{if $0\le n'<s$ and $|n|\ge s$;}\\
\frac{1}{\sqrt2}
\begin{pmatrix}
F_{|n'|n}(\mathbf q) & 0 \\
\text{sgn}(n')F_{|n'|-s,n}(\mathbf q) & 0
\end{pmatrix},
&\text{if $0\le n<s$ and $|n'|\ge s$;}\\
\frac{1}{2}
\begin{pmatrix}
F_{|n'||n|}(\mathbf q) & \text{sgn}(n)F_{|n'|,|n|-s}(\mathbf q) \\
\text{sgn}(n')F_{|n'|-s,|n|}(\mathbf q) & \text{sgn}(n'n)F_{|n'|-s,|n|-s}(\mathbf q)
\end{pmatrix},
&\text{if $|n|,|n'|\ge s$.}
\end{array}
\right.
\end{equation}
Taking the quantum mechanical expectation value, we get
\begin{equation}
\label{rhogdcgen}
\langle\hat\rho_{\xi\sigma,\xi\sigma'}(\mathbf q)\rangle=\sum'_{n,n'}\mathbf F_{n'n}(\mathbf q)
\Delta^{nn'}_{\xi\xi'\sigma\sigma'}(\mathbf q),
\end{equation}
where the guiding center density now also depends on the valley quantum numbers, i.e.,
\begin{equation}
\Delta^{nn'}_{\xi\xi'\sigma\sigma'}(\mathbf q)=\int dp e^{-iq_xp\ell^2}\langle\hat c^\dag_{n',p-q_y/2,\xi'\sigma'}\hat c_{n,p+q_y/2,\xi\sigma}\rangle.
\end{equation}

Just like before, let us approach the infinite systems as a limit of finite translation-invariant systems.
Assume
\begin{equation}
\langle\hat\rho_{\xi\sigma,\xi'\sigma'}(\mathbf q)\rangle=
\begin{pmatrix}
\rho_{11} & \rho_{12} \\
\rho_{21} & \rho_{22}
\end{pmatrix}
\frac{e^{-\frac{q^2}{4\epsilon}}}{4\pi\epsilon},\quad\quad
\Delta^{nn'}_{\xi\xi'\sigma\sigma'}(\mathbf q)=\overline\Delta^{nn'}_{\xi\xi'\sigma\sigma'}\frac{e^{-\frac{q^2}{4\epsilon}}}{4\pi\epsilon},
\end{equation}
where $\rho_{ij}$ and $\overline\Delta^{nn'}_{\xi\xi'\sigma\sigma'}$ are constants.
Plug these into Eq.~(\ref{rhogdcgen}), multiply by $e^{i\theta k}$, and integrate over the circle
$\mathcal C=\langle Q\cos\theta,Q\sin\theta\rangle$:
\begin{multline}
\delta_{k,0}
\begin{pmatrix}
\rho_{11} & \rho_{12} \\
\rho_{21} & \rho_{22}
\end{pmatrix}=
\sum_{n,n'=0}^{s-1}
\overline\Delta^{nn'}_{\xi\xi'\sigma\sigma'}
\begin{pmatrix}
\delta_{n-n',k}F_{n'n}(\mathbf Q) & 0 \\
0 & 0
\end{pmatrix}
+\sum_{n=0}^{s-1}\sum_{|n'|\ge s}
\frac{\overline\Delta^{nn'}_{\xi\xi'\sigma\sigma'}}{\sqrt2}
\begin{pmatrix}
\delta_{n-|n'|,k}F_{|n'|,n}(\mathbf Q) & 0 \\
\text{sgn}(n')\delta_{n-|n'|+s,k}F_{|n'|-s,n}(\mathbf Q) & 0
\end{pmatrix}+\\
+\sum_{n'=0}^{s-1}\sum_{|n|\ge s}
\frac{\overline\Delta^{nn'}_{\xi\xi'\sigma\sigma'}}{\sqrt2}
\begin{pmatrix}
\delta_{|n|-n',k}F_{n',|n|}(\mathbf Q) & \text{sgn}(n)\delta_{|n|-n'-s,k}F_{n',|n|-s}(\mathbf Q) \\
0 & 0
\end{pmatrix}+\\
+\sum_{|n'|,|n|\ge s}
\frac{\overline\Delta^{nn'}_{\xi\xi'\sigma\sigma'}}{2}
\begin{pmatrix}
\delta_{|n|-|n'|,k}F_{|n'|,|n|}(\mathbf Q) & \text{sgn}(n)\delta_{|n|-|n'|-s,k}F_{|n'|,|n|-s}(\mathbf Q) \\
\text{sgn}(n')\delta_{|n|-|n'|+s,k}F_{|n'|-s,|n|}(\mathbf Q) & \text{sgn}(n'n)\delta_{|n|-|n'|,k}F_{|n'|-s,|n|-s}(\mathbf Q)
\end{pmatrix}=\begin{pmatrix}
A_{11} & A_{12} \\
A_{21} & A_{22}
\end{pmatrix},
\end{multline}
where
\begin{multline}
A_{11}=\sum_{n'=0}^{s-1}\overline\Delta^{n'+k,n'}_{\xi\xi'\sigma\sigma'}F_{n',n'+k}(\mathbf Q)T(0\le n'+k<s)
+\sum_{n'=s}^{\infty}\frac{\overline\Delta^{n'+k,n'}_{\xi\xi'\sigma\sigma'}-\overline\Delta^{n'+k,-n'}_{\xi\xi'\sigma\sigma'}
}{\sqrt2}
F_{n',n'+k}(\mathbf Q)T(0\le n'+k<s)+\\
+\sum_{n'=0}^{s-1}\frac{\overline\Delta^{|n'+k|,n'}_{\xi\xi'\sigma\sigma'}+\overline\Delta^{-|n'+k|,n'}_{\xi\xi'\sigma\sigma'}
}{\sqrt2}
F_{n',|n'+k|}(\mathbf Q)T(|n'+k|\ge s)+\\
+\sum_{n'=s}^{\infty}\frac{
\overline\Delta^{|n'+k|,n'}_{\xi\xi'\sigma\sigma'}+\overline\Delta^{-|n'+k|,n'}_{\xi\xi'\sigma\sigma'}+
\overline\Delta^{|n'+k|,-n'}_{\xi\xi'\sigma\sigma'}+\overline\Delta^{-|n'+k|,-n'}_{\xi\xi'\sigma\sigma'}
}{2}
F_{n',|n'+k|}(\mathbf Q)T(|n'+k|\ge s),
\end{multline}
where $T(\dots)$ is denotes a function that yields $1$ if the logical condition in the paretheses is true,
else it yields $0$. Further,
\begin{multline}
A_{12}=
\sum_{n'=0}^{s-1}\frac{\overline\Delta^{|n'+k+s|,n'}_{\xi\xi'\sigma\sigma'}-\overline\Delta^{-|n'+k+s|,n'}_{\xi\xi'\sigma\sigma'}
}{\sqrt2}
F_{n',|n'+k+s|}(\mathbf Q)T(|n'+k+s|\ge s)+\\
+\sum_{n'=s}^{\infty}\frac{
\overline\Delta^{|n'+k+s|,n'}_{\xi\xi'\sigma\sigma'}-\overline\Delta^{-|n'+k+s|,n'}_{\xi\xi'\sigma\sigma'}
}{2}F_{n',|n'+k+s|}(\mathbf Q)T(|n'+k+s|\ge s),
\end{multline}
\begin{multline}
A_{21}=
\sum_{n'=s}^{\infty}\frac{\overline\Delta^{n'+k-s,n'}_{\xi\xi'\sigma\sigma'}-\overline\Delta^{n'+k-s,-n'}_{\xi\xi'\sigma\sigma'}
}{\sqrt2}
F_{n',n'+k-s}(\mathbf Q)T(0\le n'+k-s<s)+\\
+\sum_{n'=s}^{\infty}\frac{
\overline\Delta^{|n'+k-s|,n'}_{\xi\xi'\sigma\sigma'}-\overline\Delta^{|n'+k-s|,-n'}_{\xi\xi'\sigma\sigma'}
}{2}F_{n',|n'+k-s|}(\mathbf Q)T(|n'+k+s|\ge s),
\end{multline}
and
\begin{equation}
A_{22}=
\sum_{n'=s}^{\infty}\frac{
\overline\Delta^{|n'+k|,n'}_{\xi\xi'\sigma\sigma'}-\overline\Delta^{|n'+k|,-n'}_{\xi\xi'\sigma\sigma'}-
\overline\Delta^{-|n'+k|,n'}_{\xi\xi'\sigma\sigma'}-\overline\Delta^{-|n'+k|,-n'}_{\xi\xi'\sigma\sigma'}
}{2}
\times F_{n'-s,|n'+k|-s|}(\mathbf Q)T(|n'+k|\ge s).
\end{equation}
Now, $A_{11}=\rho_{11}\delta_{k,0}$ for and $Q$. Then the coefficient of any $F_{n',|n'+k|}(\mathbf Q)$
must vanish for all $n'\ge 0$. This yields
\begin{gather}
\overline\Delta^{n'+k,n'}_{\xi\xi'\sigma\sigma'}\propto\delta_{k,0}\quad\text{if $0\le n',n'+k<s$},\label{cond1}\\
\left(\overline\Delta^{|n'+k|,n'}_{\xi\xi'\sigma\sigma'}+\overline\Delta^{-|n'+k|,n'}_{\xi\xi'\sigma\sigma'}\right)
\propto\delta_{k,0}\quad\text{if $0\le n'<s$ and $|n'+k|\ge s$},\label{cond2}\\
\left(\overline\Delta^{n'+k,n'}_{\xi\xi'\sigma\sigma'}+\overline\Delta^{n'+k,-n'}_{\xi\xi'\sigma\sigma'}\right)
\propto\delta_{k,0}\quad\text{if $0\le n'+k<s$ and $n'\ge s$},\label{cond3}\\
\left(\overline\Delta^{|n'+k|,n'}_{\xi\xi'\sigma\sigma'}+\overline\Delta^{|n'+k|,-n'}_{\xi\xi'\sigma\sigma'}+
\overline\Delta^{-|n'+k|,n'}_{\xi\xi'\sigma\sigma'}+\overline\Delta^{-|n'+k|,-n'}_{\xi\xi'\sigma\sigma'}
\right)
\propto\delta_{k,0}\quad\text{if $|n'+k|,n'\ge s$}.\label{cond4}
\end{gather}
Similarly, from the condition $A_{12}=\rho_{11}\delta_{k,0}$ for and $Q$ we get
\begin{equation}
\left(\overline\Delta^{|n'+k+s|,n'}_{\xi\xi'\sigma\sigma'}-\overline\Delta^{-|n'+k+s|,n'}_{\xi\xi'\sigma\sigma'}\right)
\propto\delta_{k,0}\quad\text{if $|n'+k+s|\ge s$}.\label{cond5}
\end{equation}
From the condition $A_{21}=\rho_{21}\delta_{k,0}$ for and $Q$ we get
\begin{equation}
\left(\overline\Delta^{|n'+k-s|,n'}_{\xi\xi'\sigma\sigma'}-\overline\Delta^{|n'+k-s|,-n'}_{\xi\xi'\sigma\sigma'}\right)
\propto\delta_{k,0}\quad\text{if $n'\ge s$}.\label{cond6}
\end{equation}
Finally, for $n',|n'+k|\ge s$ the condition $A_{22}=\rho_{22}\delta_{k,0}$ for and $Q$ yields
\begin{equation}
\left(\overline\Delta^{|n'+k|,n'}_{\xi\xi'\sigma\sigma'}-\overline\Delta^{|n'+k|,-n'}_{\xi\xi'\sigma\sigma'}-
\overline\Delta^{-|n'+k|,n'}_{\xi\xi'\sigma\sigma'}+\overline\Delta^{-|n'+k|,-n'}_{\xi\xi'\sigma\sigma'}
\right)
\propto\delta_{k,0}.\label{cond7}
\end{equation}
Now, Eq.~(\ref{cond1}) implies $\overline\Delta^{m,m'}_{\xi\xi'\sigma\sigma'}\propto\delta_{m,m'}$ for $0\le m,m'<s$.
Eqs.~(\ref{cond2}) and (\ref{cond5}) imply $\overline\Delta^{m,m'}_{\xi\xi'\sigma\sigma'}\propto\delta_{|m|,m'}$
for $0\le m'<s$ and $|m|>s$.
Eqs.~(\ref{cond3}) and (\ref{cond6}) imply $\overline\Delta^{m,m'}_{\xi\xi'\sigma\sigma'}\propto\delta_{m,|m'|}$
for $0\le m<s$ and $|m'|>s$.

Further, fix some $m,m'\ge s$.
Subtracting Eqs.~(\ref{cond5}) and (\ref{cond6}) imply
\begin{equation}
\left(\overline\Delta^{m,-m'}_{\xi\xi'\sigma\sigma'}-\overline\Delta^{-m,m'}_{\xi\xi'\sigma\sigma'}\right)
\propto\delta_{m,m'}.\label{cons1}
\end{equation}
Subtracting Eqs.~(\ref{cond4}) and (\ref{cond7}) imply
\begin{equation}
\left(\overline\Delta^{m,-m'}_{\xi\xi'\sigma\sigma'}+\overline\Delta^{-m,m'}_{\xi\xi'\sigma\sigma'}\right)
\propto\delta_{m,m'}.\label{cons2}
\end{equation}
Eqs.~(\ref{cons1}) and (\ref{cons2}) imply $\overline\Delta^{m,-m'}_{\xi\xi'\sigma\sigma'}\propto\delta_{m,m'}$
and $\overline\Delta^{-m,m'}_{\xi\xi'\sigma\sigma'}\propto\delta_{m,m'}$.
The using Eq.~(\ref{cond5}) we obtain $\overline\Delta^{m,m}_{\xi\xi'\sigma\sigma'}\propto\delta_{m,m'}$.
Finally, from Eq.~(\ref{cond5}) we get $\overline\Delta^{-m,-m}_{\xi\xi'\sigma\sigma'}\propto\delta_{m,m'}$.
To summarize,
\begin{equation}
\overline\Delta^{m,m'}_{\xi\xi'\sigma\sigma'}\propto\delta_{|m|,|m'|}.
\end{equation}

The mean field Hamiltonian can be written in terms of the guiding center density; the derivation is
identical to the steps between Eqs.~(\ref{gdcfrom}) and (\ref{gdcto}), apart from the more cumbersome
notation. We omit this trivially reproducible part.

This completes the argument that the summation in Eq.~(4) of the paper can be restricted to $|m|=|m'|$ for
homogeneous liquid states, e.g., for integer quantum Hall states.

\section{The integer quantum Hall ground state at $\nu=\pm2s$ in the mean-field approximation}
\label{seciqhe}

At $\nu=-2s$ the zero-energy Landau band is empty and at $\nu=2s$ it is completely filled;
even single-particle considerations imply a gap; this is the integer quantum Hall effect.

We consider the Ansatz
\begin{equation}\label{igheqnsatz}
\Psi^{(s)}_0=\prod_{m=s}^\infty\prod_q\prod_{\xi,\sigma}
\left(
\cos\left(\frac{\theta_{m\xi\sigma}}{2}\right)e^{i\phi_{m\xi\sigma}/2}\hat c^\dag_{-mq\xi\sigma}
+\sin\left(\frac{\theta_{m\xi\sigma}}{2}\right)e^{-i\phi_{m\xi\sigma}/2}\hat c^\dag_{mq\xi\sigma}
\right)\left|0\right\rangle,
\end{equation}
where $\theta_{m\xi\sigma}$ and $\phi_{m\xi\sigma}$ are parameters for $m\ge s$.
The single-particle Hamiltonian is
\begin{equation}
\hat H_0=\varepsilon_s
\sum_{m=s}^\infty\sqrt{\prod_{i=0}^{s-1}(m-i)}\sum_{q\xi\sigma}
\left(\hat c_{mq\xi\sigma}^\dag\hat c_{mq\xi\sigma} - \hat c_{-mq\xi\sigma}^\dag\hat c_{-mq\xi\sigma}\right).
\end{equation}
The expectation value of the single-particle Hamiltonian in state $\Psi^{(s)}_0$ is
\begin{multline}
\langle\Psi^{(s)}_0|\hat H_0|\Psi^{(s)}_0\rangle=\varepsilon_s
\sum_{m=s}^\infty
\sqrt{\prod_{i=0}^{s-1}(m-i)}\sum_{q\xi\sigma}
\left(
\sin^2\left(\frac{\theta_{m\xi\sigma}}{2}\right) -
\cos^2\left(\frac{\theta_{m\xi\sigma}}{2}\right)
\right)=\\
=-N_\phi\varepsilon_s
\sum_{m=s}^\infty\sqrt{\prod_{i=0}^{s-1}(m-i)}\sum_{\xi\sigma}\cos\theta_{m\xi\sigma}.
\end{multline}
In itself, $\hat H_0$ would be minimized by $\theta_{m\xi\sigma}=0$ for all $m$.

The interaction part of the Hamiltonian is considered in the Hartree-Fock mean-field approximation, c.f.\ Eq.~(\ref{meanfield2}).
The relevant expectation values are
\begin{equation}
\left\langle \hat c^\dag_{m'p}\hat c_{mp}\right\rangle=
\left\{
\begin{array}{ll}
\sin^2\left(\frac{\theta_{m}}{2}\right) & \text{if $m=m'\ge s$,}\\
\cos^2\left(\frac{\theta_{|m|}}{2}\right) & \text{if $m=m'\le-s$,}\\
\sin\left(\frac{\theta_{m}}{2}\right)\cos\left(\frac{\theta_{m}}{2}\right)e^{-i\phi_m} & \text{if $m=-m'\ge s$,}\\
\sin\left(\frac{\theta_{|m|}}{2}\right)\cos\left(\frac{\theta_{|m|}}{2}\right)e^{i\phi_{|m|}} & \text{if $m=-m'\le-s$.}
\end{array}
\right.
\end{equation}
Thus the mean-field ground state energy is, suppressing spin and valley for simplicity,
\begin{align*}
\frac{E^{(s)}_0}{N_\phi}&=
\langle\Psi^{(s)}_0|\frac{\hat H_0+\hat V^\text{F}}{N_\phi}|\Psi^{(s)}_0\rangle=\\
&-\frac{1}{2}\sum_{|n|=|n'|}\sum_{|m|=|m'|}\int\frac{d^2k}{(2\pi)^2}F^{(s)}_{nm}(\mathbf k)F^{(s)}_{m'n'}(-\mathbf k)\frac{e^2}{4\pi\epsilon}\frac{2\pi}{k}
\left\langle \hat c^\dag_{m'p}\hat c_{mp}\right\rangle
\left\langle\hat c^\dag_{n,p-k_y}\hat c_{n'_,p-k_y}\right\rangle=\\
&-\varepsilon_s\sum_{m=s}^\infty\sqrt{\prod_{i=0}^{s-1}(m-i)}\cos\theta_m
-\frac{1}{2}\sum_{n=s}^\infty\sum_{m=s}^\infty\int\frac{d^2k}{(2\pi)^2}\frac{e^2}{4\pi\epsilon}\frac{2\pi}{k}
\left[F^{(s)}_{nm}(\mathbf k)F^{(s)}_{mn}(-\mathbf k)\sin^2\left(\frac{\theta_{m}}{2}\right)\sin^2\left(\frac{\theta_{n}}{2}\right)\right.+\\
&F^{(s)}_{n,-m}(\mathbf k)F^{(s)}_{-mn}(-\mathbf k)\cos^2\left(\frac{\theta_{m}}{2}\right)\sin^2\left(\frac{\theta_{n}}{2}\right)
+F^{(s)}_{-nm}(\mathbf k)F^{(s)}_{m,-n}(-\mathbf k)\sin^2\left(\frac{\theta_{m}}{2}\right)\cos^2\left(\frac{\theta_{n}}{2}\right)+\\
&F^{(s)}_{-n,-m}(\mathbf k)F^{(s)}_{-m,-n}(-\mathbf k)\cos^2\left(\frac{\theta_{m}}{2}\right)\cos^2\left(\frac{\theta_{n}}{2}\right)+\\
&F^{(s)}_{nm}(\mathbf k)F^{(s)}_{m,-n}(-\mathbf k)\sin^2\left(\frac{\theta_{m}}{2}\right)
\sin\left(\frac{\theta_{n}}{2}\right)\cos\left(\frac{\theta_{n}}{2}\right)\left(e^{i\phi_n}+e^{-i\phi_n}\right)+\\
&F^{(s)}_{n,-m}(\mathbf k)F^{(s)}_{-m,-n}(-\mathbf k)\cos^2\left(\frac{\theta_{m}}{2}\right)
\sin\left(\frac{\theta_{n}}{2}\right)\cos\left(\frac{\theta_{n}}{2}\right)\left(e^{i\phi_n}+e^{-i\phi_n}\right)+\\
&F^{(s)}_{nm}(\mathbf k)F^{(s)}_{-m,n}(-\mathbf k)\sin^2\left(\frac{\theta_{n}}{2}\right)
\sin\left(\frac{\theta_{m}}{2}\right)\cos\left(\frac{\theta_{m}}{2}\right)\left(e^{i\phi_m}+e^{-i\phi_m}\right)+\\
&F^{(s)}_{-nm}(\mathbf k)F^{(s)}_{-m,-n}(-\mathbf k)\cos^2\left(\frac{\theta_{n}}{2}\right)
\sin\left(\frac{\theta_{m}}{2}\right)\cos\left(\frac{\theta_{m}}{2}\right)\left(e^{i\phi_m}+e^{-i\phi_m}\right)+\\
&F^{(s)}_{nm}(\mathbf k)F^{(s)}_{-m,-n}(-\mathbf k)\sin\left(\frac{\theta_{m}}{2}\right)\cos\left(\frac{\theta_{m}}{2}\right)
\sin\left(\frac{\theta_{n}}{2}\right)\cos\left(\frac{\theta_{n}}{2}\right)
\left(e^{i(\phi_m+\phi_n)}+e^{-i(\phi_m+\phi_n)}\right)+\\
&\left.F^{(s)}_{n,-m}(\mathbf k)F^{(s)}_{m,-n}(-\mathbf k)\sin\left(\frac{\theta_{m}}{2}\right)\cos\left(\frac{\theta_{m}}{2}\right)
\sin\left(\frac{\theta_{n}}{2}\right)\cos\left(\frac{\theta_{n}}{2}\right)
\left(e^{i(\phi_m-\phi_n)}+e^{-i(\phi_m-\phi_n)}\right)\right]
\end{align*}
Then
\begin{multline}
\frac{E^{(s)}_0}{N_\phi}=
-\varepsilon_s\sum_{m=s}^\infty\sqrt{\prod_{i=0}^{s-1}(m-i)}\cos\theta_{m}
-\frac{1}{2}\sum_{n=s}^\infty\sum_{m=s}^\infty
\left[X_{nm}w^+(\theta_m,\phi_m)w^+(\theta_n,\phi_n)+X_{n-s,m-s}w^-(\theta_m,\phi_m)w^-(\theta_n,\phi_n)\right.+\\
\left.+\frac{1}{2}X^{(s)}_{nm}\left(\cos\theta_m\cos\theta_n-\sin\theta_m\sin\phi_m\sin\theta_n\sin\phi_n\right)\right],
\end{multline}
where I have defined
\begin{equation}
\label{distri}
w^\pm(\theta,\phi)=\frac{1\pm\sin\theta\cos\phi}{2}.
\end{equation}
This quantity has a very simple meaning.
The parametrization in Eq.~(\ref{igheqnsatz}) lets the orbitals in the two-dimensional subspace with fixed $|m|$ rotate.
The $\Psi_{mq}$ and $\Psi_{-mq}$ orbitals differ only by a relative phase between the top and bottom spinor components.
Thus the linear combination  in Eq.~(\ref{igheqnsatz}) changes the probability of finding the electron in the two sublattices that correspond to these spinor components.
In the filled state the electron will have weight $w^+(\theta,\phi)$ in the sublattice that corresponds to the top component,
$w^-(\theta,\phi)$ in the other one; the converse holds for the state that is left empty in state $\Psi^{(s)}_0$.

Using the identity, to be proved in Section \ref{identities},
\begin{equation}
\label{iden}
\sum_n|F_{n'n}(\mathbf k)|^2=1
\end{equation}
for any fixed $n'$, we obtain
\begin{equation}
\label{regularization}
\sum_{m=M+1}^\infty X_{nm}=-\sum_{m=0}^{M} X_{nm}+\frac{e^2}{4\pi\epsilon}\int\frac{d^2k}{(2\pi)^2}\frac{2\pi}{k}.
\end{equation}
The last term is infinite in the thermodynamical limit,
but irrelevant from the point of view of energy minimization if it has no parameter-dependent prefactor.
We drop such infinite constants from any energy expression.
Then
\begin{multline}
\label{emptyenergy}
\frac{E^{(s)}_0}{N_\phi}=
-\varepsilon_s\sum_{m=s}^\infty\sqrt{\prod_{i=0}^{s-1}(m-i)}\cos\theta_{m}-\\
-\frac{1}{8}\sum_{n=s}^\infty\sum_{m=s}^\infty
\left[(X_{nm}-X_{n-s,m-s})\sin\theta_m\cos\phi_m\sin\theta_n\cos\phi_n
+2X^{(s)}_{nm}\left(\cos\theta_m\cos\theta_n-\sin\theta_m\sin\phi_m\sin\theta_n\sin\phi_n\right)\right]-\\
-\frac{1}{8}\sum_{n=s}^\infty\sum_{m=s}^\infty(X_{nm}+X_{n-s,m-s})+
\frac{1}{4}\sum_{n=0}^{s-1}\sum_{m=s}^\infty X_{nm}\sin\theta_m\cos\phi_m.
\end{multline}
It is instructive to see how particle-hole symmetry is manifest in this simple case.
Consider the state in which the zero-energy Landau band is completely filled,
\begin{equation}
\Psi^{(s)}_{s}=\left(\prod_q\prod_{n=0}^{s-1}\hat c^\dag_{nq}\right)\Psi^{(s)}_0.
\end{equation}
The expectation values in this state are
\begin{equation}
\left\langle \hat c^\dag_{m'p}\hat c_{mp}\right\rangle=
\left\{
\begin{array}{ll}
\sin^2\left(\frac{\theta_{m}}{2}\right) & \text{if $m=m'\ge s$,}\\
\cos^2\left(\frac{\theta_{|m|}}{2}\right) & \text{if $m=m'\le-s$,}\\
\sin\left(\frac{\theta_{m}}{2}\right)\cos\left(\frac{\theta_{m}}{2}\right)e^{-i\phi_m} & \text{if $m=-m'\ge s$,}\\
\sin\left(\frac{\theta_{|m|}}{2}\right)\cos\left(\frac{\theta_{|m|}}{2}\right)e^{i\phi_{|m|}} & \text{if $m=-m'\le-s$.}\\
1 & \text{if $0\le m=m'<s$.}
\end{array}
\right.
\end{equation}
The ground state energy per particle is
\begin{equation}
\frac{E^{(s)}_{s}}{N_\phi}=
\frac{E^{(s)}_0}{N_\phi}+\frac{1}{2}\sum_{n=0}^{s-1}\sum_{m=s}^\infty X_{nm}\sin\theta_m\cos\phi_m.
\end{equation}
That is, in comparison to the energy of the state $\Psi^{(s)}_0$ at filling factor $\nu=-2s$, the sign of the last term in Eq.~(\ref{emptyenergy}) changes.
This is the only term that contains the product of an odd number of sine/cosine functions of the phase $\phi_m$.
Thus if $\{\theta_m,\phi_m\}$ optimizes the ground state energy at $\nu=-2s$,
$\{\theta_m,\phi_m+\pi\}$ optimizes the ground state energy at $\nu=2s$.

Let us introduce a Landau level cutoff at $-M$, i.e., let us set $\theta_m=0$ for all $m>M$. Then
\begin{multline}
\frac{E^{(s)}_0}{N_\phi}=
-\varepsilon_s\sum_{m=s}^\infty\sqrt{\prod_{i=0}^{s-1}(m-i)}\cos\theta_{m}
-\frac{1}{2}\sum_{n=s}^M\sum_{m=s}^M
\left[X_{nm}w^+(\theta_m,\phi_m)w^+(\theta_n,\phi_n)+X_{n-s,m-s}w^-(\theta_m,\phi_m)w^-(\theta_n,\phi_n)+\right.\\
\left.+\frac{1}{2}X^{(s)}_{nm}\left(\cos\theta_m\cos\theta_n-\sin\theta_m\sin\phi_m\sin\theta_n\sin\phi_n\right)\right]-
\frac{1}{8}\sum_{n=M+1}^\infty\sum_{m=M+1}^\infty\left[X_{nm}+X_{n-s,m-s}+2X^{(s)}_{nm}\right]-\\
-\frac{1}{4}\sum_{n=s}^M\sum_{m=M+1}^\infty
\left[X_{nm}w^+(\theta_n,\phi_n)+X_{n-s,m-s}w^-(\theta_n,\phi_n)+X^{(s)}_{nm}\cos\theta_n\right]-\\
-\frac{1}{4}\sum_{n=M+1}^\infty\sum_{m=s}^s
\left[X_{nm}w^+(\theta_m,\phi_m)+X_{n-s,m-s}w^-(\theta_m,\phi_m)+X^{(s)}_{nm}\cos\theta_m\right].
\end{multline}
Using Eq.~(\ref{regularization}), elementary algebra yields
\begin{multline}
\frac{E^{(s)}_0}{N_\phi}=
-\varepsilon_s\sum_{m=s}^\infty\sqrt{\prod_{i=0}^{s-1}(m-i)}\cos\theta_{m}
-\frac{1}{8}\sum_{n=s}^M\sum_{m=s}^M
\left(X_{nm} + X_{n-s,m-s}\right)\sin\theta_m\cos\phi_m\sin\theta_n\cos\phi_n-\\
-\frac{1}{4}\sum_{n=s}^M\sum_{m=s}^M
X^{(s)}_{nm}\left(\cos\theta_m\cos\theta_n-\sin\theta_m\sin\phi_m\sin\theta_n\sin\phi_n\right)
-\frac{1}{2}\sum_{n=s}^MA_n\cos\theta_n
+\frac{1}{4}\sum_{n=s}^MB_n\sin\theta_n\cos\phi_n,
\end{multline}
where
\begin{equation}
A_n=\sum_{m=M+1}^\infty X^{(s)}_{nm},\quad\text{and}\quad
B_n=\sum_{m=0}^{s-1} X_{nm}.
\end{equation}
It can be checked that $A_n$ is finite.
The minimum is found numerically using a quasi-Newton method.
Fig.~1 shows the charge distribution in Eq.~(\ref{distri}) for the optimized parameters.
With increasing $|m|$ the mean-field ground state has a rapidly increasing weight on the negative-energy single particle orbital.
That is, the interaction rotates only the states in the topmost Landau levels, which justifies \textit{a posteriori} our cutoff scheme.
In the figures we used a cutoff $M=12$; we have checked, however, that a larger cutoff hardly changes the results.

\section{The quantum Hall ferromagnets at in the zero-energy Landau band of bilayer graphene}

For the clarity of the presentation, we first focus on the orbital structure in the zero-energy Landau
band, ignoring the spin and valley pseudospin degrees of freedom.
The inclusion of these degrees of freedom is easy, and it is presented in Subsection \ref{spin}.
We will follow the same procedure for multilayers in subsequent sections.

We introduce the relative filling factor in the zero-energy Landau band as
\begin{equation}
\overline\nu=\nu+4
\end{equation}
for the purposes of this section.
In the section on trilayer and four-layer graphene, $\overline\nu$ will be redefined appropriately.

\subsection{Suppressing spin and valley}
\label{spinless}

We seek the mean-field ground state at relative filling factor $\overline\nu=1$ of the zero-energy Landau band of bilayer graphene
in the form
\begin{equation}
\Psi^{(2)}_{1}=\prod_q
\left(\cos\left(\frac{\theta}{2}\right)e^{i\phi/2}\hat c^\dag_{0q}+\sin\left(\frac{\theta}{2}\right)e^{-i\phi/2}\hat c^\dag_{1q}
\right)\Psi^{(2)}_0,
\end{equation}
where $\Psi^{(2)}_0$ is the generic form of the integer quantum Hall state at $\nu=-4$ defined in Eq.~(\ref{igheqnsatz}).
In the paper $z_0=\cos\left(\frac{\theta}{2}\right)e^{i\phi/2}$ and $z_1=\sin\left(\frac{\theta}{2}\right)e^{-i\phi/2}$ is used for brevity.
The relevant expectation values are
\begin{equation}
\left\langle \hat c^\dag_{m'p}\hat c_{mp}\right\rangle=
\left\{
\begin{array}{ll}
\sin^2\left(\frac{\theta_{m}}{2}\right) & \text{if $m=m'\ge s$,}\\
\cos^2\left(\frac{\theta_{|m|}}{2}\right) & \text{if $m=m'\le-s$,}\\
\sin\left(\frac{\theta_{m}}{2}\right)\cos\left(\frac{\theta_{m}}{2}\right)e^{-i\phi_m} & \text{if $m=-m'\ge s$,}\\
\sin\left(\frac{\theta_{|m|}}{2}\right)\cos\left(\frac{\theta_{|m|}}{2}\right)e^{i\phi_{|m|}} & \text{if $m=-m'\le-s$,}\\
\cos^2\left(\frac{\theta}{2}\right) & \text{if $m=m'=0$,}\\
\sin^2\left(\frac{\theta}{2}\right) & \text{if $m=m'=1$,}\\
\sin\left(\frac{\theta}{2}\right)\cos\left(\frac{\theta}{2}\right)e^{i\phi} & \text{if $m=0,m'=1$,}\\
\sin\left(\frac{\theta}{2}\right)\cos\left(\frac{\theta}{2}\right)e^{-i\phi} & \text{if $m=1,m'=0$.}
\end{array}
\right.
\end{equation}
Then
\begin{multline}
\label{bolgen2}
\frac{E^{(2)}_{1}}{N_\phi}=\frac{E^{(2)}_0}{N_\phi}
-\frac{X_{00}}{2}\cos^4\left(\frac{\theta}{2}\right)-\frac{X_{11}}{2}\sin^4\left(\frac{\theta}{2}\right)
-(X_{01}+X_{0011})\sin^2\left(\frac{\theta}{2}\right)\cos^2\left(\frac{\theta}{2}\right)+\\
-\sum_{m=2}^\infty\left(\cos^2\left(\frac{\theta}{2}\right) X_{0m}+\sin^2\left(\frac{\theta}{2}\right) X_{1m}\right)w^+(\theta_m,\phi_m).
\end{multline}
For the values of the exchange integral constants see Table \ref{xtable}.
Notice that $E^{(2)}_0$ contains the exchange interaction of the $|m|\ge2$ Landau levels and the total kinetic energy.
Clearly, the first line of Eq.~(\ref{bolgen2}) contains the exchange interaction within the zero-energy Landau band,
while the second line is the exchange between the states in the zero-energy Landau band and the deeper-lying filled states.
As the states in the zero-energy Landau band are contained in the sublattice that corresponds to the top spinor component, only the
weight $w^+(\theta_m,\phi_m)$ occurs.

\begin{table}[htb]
\begin{center}
\begin{tabular}{c|c|c|c|c}
\hline\hline
$X_{nm}$ & $n=0$ & $n=1$ & $n=2$ & $n=3$\\
\hline
$m=0$ & $\sqrt\frac{\pi}{2}$ & $\frac{1}{2}\sqrt\frac{\pi}{2}$ & $\frac{3}{8}\sqrt\frac{\pi}{2}$ & $\frac{5}{16}\sqrt\frac{\pi}{2}$ \\
$m=1$ & $\frac{1}{2}\sqrt\frac{\pi}{2}$ & $\frac{3}{4}\sqrt\frac{\pi}{2}$ & $\frac{7}{16}\sqrt\frac{\pi}{2}$ & $\frac{11}{32}\sqrt\frac{\pi}{2}$ \\
$m=2$ & $\frac{3}{8}\sqrt\frac{\pi}{2}$ & $\frac{7}{16}\sqrt\frac{\pi}{2}$ & $\frac{41}{64}\sqrt\frac{\pi}{2}$ & $\frac{51}{128}\sqrt\frac{\pi}{2}$ \\
$m=3$ & $\frac{5}{16}\sqrt\frac{\pi}{2}$ & $\frac{11}{32}\sqrt\frac{\pi}{2}$ & $\frac{51}{128}\sqrt\frac{\pi}{2}$ & $\frac{147}{256}\sqrt\frac{\pi}{2}$ \\
\hline\hline
\end{tabular}
\end{center}
\caption{\label{xtable}
Exchange integral constants defined in Eq.~(\ref{exchange}).
Notice $X_{iijj}=X_{ij}$, where $X_{iijj}$ was defined in Eq.~(\ref{exint}).
The units are $\frac{e^2}{4\pi\epsilon\ell}$.
}
\end{table}

As $\frac{\partial}{\partial\phi}E^{(2)}_{1}=0$, the ground state manifold has U(1) symmetry.
The actual ground state will break this symmetry, and there is an orbital Goldstone mode associated with it.

In the weak-coupling limit, i.e., $\beta^{(2)}\ll1$, $\theta_m=0$ and the $\phi_m$'s are irrelevant.
$\theta$ is then determined by the the minimization of the terms beyond $E^{(2)}_0$:
\begin{multline}
\frac{E^{(2)}_{1}-E^{(2)}_0}{N_\phi}=
-\frac{1}{2}\left[X_{00}\cos^4\left(\frac{\theta}{2}\right)+ X_{11}\sin^4\left(\frac{\theta}{2}\right) +
2(X_{01}+X_{0011})\sin^2\left(\frac{\theta}{2}\right)\cos^2\left(\frac{\theta}{2}\right)\right.+\\
\left.+\sum_{m=2}^\infty
\left(\cos^2\left(\frac{\theta}{2}\right) X_{0m}+\sin^2\left(\frac{\theta}{2}\right) X_{1m}\right)\right].
\end{multline}
With regularization according to Eq.~(\ref{regularization}) and dropping simple constants,
\begin{multline}
\frac{E^{(2)}_{1}-E^{(2)}_0}{N_\phi}=
-\frac{1}{2}\left[X_{00}\cos^4\left(\frac{\theta}{2}\right)+X_{11}\sin^4\left(\frac{\theta}{2}\right) +\right.\\
\left.+2(X_{01}+X_{0011})\sin^2\left(\frac{\theta}{2}\right)\cos^2\left(\frac{\theta}{2}\right)
-\cos^2\left(\frac{\theta}{2}\right)X_{00}-\sin^2\left(\frac{\theta}{2}\right)X_{11}\right].\label{bilayerzeroth}
\end{multline}
Using Table \ref{xtable}, the energy expression in Eq.~(\ref{bilayerzeroth}) is minimized by $\theta=\pm\frac{\pi}{2}$
or $|z_0|=|z_1|=\frac{1}{\sqrt2}$, i.e., the mean-field ground state in the zero-energy
Landau level is an equal-weight linear combination of the $n=0$ and $n=1$ orbitals, with an arbitrary relative phase.

Consider the general case of a finite $\beta^{(2)}$.
First, one can show analytically that $\theta=\pi/2$ always remains an extremum.
Plugging $\theta=\pi/2$ into Eq.~(\ref{bolgen2}) yields
\begin{equation}
\frac{E^{(2)}_1}{N_\phi}=\frac{E^{(2)}_0}{N_\phi}+\frac{1}{4}\sum_{n=0}^{1}\sum_{m=s}^\infty X_{nm}\sin\theta_m\cos\phi_m+\text{const.}
\end{equation}
Thus, the last term in Eq.~(\ref{emptyenergy}) is cancelled, and each remaining term contains an even number of sine/cosine function of $\phi_m$'s.
Physically, this means that with each orbital state in the zero-energy Landau band filled with weight $\frac{1}{2}$;
there is no preferential sublattice.
The remaining terms are then extremized by setting all $\theta_m=0$.
States in the zero-energy Landau band with this property, namely that the exchange energy of the $|m|\ge s$ with the zero-energy Landau band exactly removes the sublattice asymmetry, i.e., the $|m|\ge s$ states can no longer lower their interaction energy
by producing a larger weight in the bottom spinorial component, will be named ``balanced'' states.
For bilayer graphene the U(1) manifold of balanced states is its own particle-hole conjugate.

However, it is unclear if $\theta=\pi/2,\theta_m=0$ is a global minimum.
We minimize $E^{(2)}_{1}$ is with a Landau level cutoff $-M$.
Setting $\theta_m=0$ for $m>M$, and using the usual regularization in Eq.~(\ref{regularization}) for infinite
summations of the $X_{nm}$ exchange constants, elementary algebra yields
\begin{multline}
\frac{E^{(2)}_{1}}{N_\phi}=\frac{E^{(2)}_0}{N_\phi}-
\frac{1}{2}\left[
X_{00}\left(\cos^4\left(\frac{\theta}{2}\right)-\cos^2\left(\frac{\theta}{2}\right)\right)+ 
X_{11}\left(\sin^4\left(\frac{\theta}{2}\right)-\sin^2\left(\frac{\theta}{2}\right)\right)+
2(X_{01}+X_{0011})\sin^2\left(\frac{\theta}{2}\right)\cos^2\left(\frac{\theta}{2}\right)\right]\\
+\frac{1}{2}\sum_{m=0}^{s-1}\left(\cos^2\left(\frac{\theta}{2}\right) X_{0m} + \sin^2\left(\frac{\theta}{2}\right) X_{1m}\right)
-\frac{1}{2}\sum_{m=s}^M\left(\cos^2\left(\frac{\theta}{2}\right) X_{0m} + \sin^2\left(\frac{\theta}{2}\right) X_{1m}\right)
\sin\theta_m\cos\phi_m.
\end{multline}
This expression can be minimized numerically.
As Fig.~2(a) demonstrates, the optimal weight of the orbitals in the zero-energy Landau band bifurcates
around $\beta^{(2)}_c\approx3.56$.
In other words, the balanced state remains stable for finite interaction strength up to a point,
but eventually yields to a pair
of states that break the balance of the zero-energy Landau band and lower the total exchange interaction energy
by mixing the $-m$ and $+m$
Landau orbitals, moving charge from the sublattice that corresponds to the upper spinor component to
the sublattice that corresponds to the lower spinor component.
While the balanced state is its own particle-hole conjugate, the pair of states for $\beta^{(2)}>\beta^{(2)}_c$
one another's particle-hole conjugate.

\subsection{Including spin and valley}
\label{spin}

The inclusion of the spin and valley pseudospin degrees of freedom is easy, because the mean-field interaction Hamiltonian
only consists of the exchange part for homogeneous states, and exchange does not couple states with different spin or valley.
Consider the Ansatz
\begin{equation}
\tilde\Psi^{(2)}_{1}=\prod_q\prod_{\xi,\sigma}
\left(z_{0\xi\sigma}\hat c^\dag_{0q\xi\sigma}+z_{1\xi\sigma}\hat c^\dag_{1q\xi\sigma}
\right)\Psi^{(2)}_0,
\end{equation}
where $\Psi^{(2)}_0$ is still the generic state at $\nu=-4$ defined in Eq.~(\ref{igheqnsatz}),
parametrized by $\{\theta_{m\xi\sigma},\phi_{m\xi\sigma}\}$.
Normalization requires
\begin{equation}
\sum_{\xi,\sigma}\left(|z_{0\xi\sigma}|^2+|z_{1\xi\sigma}|^2\right)=1.
\end{equation}
Following the same steps as before, we obtain
\begin{multline}
\frac{\tilde E^{(2)}_{1}}{N_\phi}=\frac{E^{(2)}_0}{N_\phi}-
\frac{1}{2}\sum_{\xi,\sigma}\left[
X_{00}\left(|z_{0\xi\sigma}|^4-|z_{0\xi\sigma}|^2\right)+
X_{11}\left(|z_{1\xi\sigma}|^4-|z_{1\xi\sigma}|^2\right)+2(X_{01}+X_{0011})|z_{0\xi\sigma}|^2|z_{1\xi\sigma}|^2\right]+\\
+\frac{1}{2}\sum_{\xi,\sigma}
\sum_{m=0}^{s-1}\left(|z_{0\xi\sigma}|^2 X_{0m} + |z_{1\xi\sigma}|^2 X_{1m}\right)-\frac{1}{2}\sum_{\xi,\sigma}
\sum_{m=s}^M\left(|z_{0\xi\sigma}|^2 X_{0m} + |z_{1\xi\sigma}|^2 X_{1m}\right)
\sin\theta_m\cos\phi_m.
\end{multline}
Let us introduce new variables
\begin{equation}
Z_{\xi\sigma}=|z_{0\xi\sigma}|^2+|z_{1\xi\sigma}|^2\quad\text{and}\quad
Y_{\xi\sigma}=\frac{|z_{0\xi\sigma}|^2}{Z_{\xi\sigma}}.
\end{equation}
The domain of these variables are, of course,
\begin{equation}
\label{domain}
0\le Z_{\xi\sigma}\le 1\quad\text{and}\quad 0\le Y_{\xi\sigma}\le 1.
\end{equation}
The mean-field ground state energy can be written in the convenient form parametrized by seven independent real parameters:
\begin{multline}
\label{spinfulbilayer}
\frac{\tilde E^{(2)}_{1}}{N_\phi}=\frac{E^{(2)}_0}{N_\phi}-
\sum_{\xi,\sigma}
\left[
\frac{X_{00}}{2}Z_{\xi\sigma}Y_{\xi\sigma}(1-Z_{\xi\sigma}Y_{\xi\sigma})+
\frac{X_{11}}{2}Z_{\xi\sigma}(1-Y_{\xi\sigma})\left(1-Z_{\xi\sigma}(1-Y_{\xi\sigma})\right)
-(X_{01}+X_{0011})Z_{\xi\sigma}^2Y_{\xi\sigma}(1-Y_{\xi\sigma})\right]\\
+\frac{1}{2}\sum_{\xi,\sigma}\sum_{m=0}^{s-1}\left(Z_{\xi\sigma}Y_{\xi\sigma}X_{0m} + Z_{\xi\sigma}(1-Y_{\xi\sigma})X_{1m}\right)
-\frac{1}{2}\sum_{\xi,\sigma}\sum_{m=s}^M\left(Z_{\xi\sigma}Y_{\xi\sigma}X_{0m} + Z_{\xi\sigma}(1-Y_{\xi\sigma})X_{1m}\right)\sin\theta_m\cos\phi_m.
\end{multline}
The global minimum of the function $\tilde E^{(2)}_{1}$ in Eq.~(\ref{spinfulbilayer}) is outside of the domain in Eq.~(\ref{domain}).
Hence we minimize $\tilde E^{(2)}_{1}$ on the boundary of the domain.
For $\beta^{(2)}<\beta^{(2)}_c\approx3.56$ there are two local minima apart from the permutation of the spin/valley components,
\begin{gather}
\{Z_{\xi\sigma}\}=\{1,0,0,0\}, \quad \{Y_{\xi\sigma}\}=\{\frac{1}{2},-,-,-\},\label{global}\\
\{Z_{\xi\sigma}\}=\{\frac{1}{2},\frac{1}{2},0,0\}, \quad \{Y_{\xi\sigma}\}=\{0,0,-,-\}.
\end{gather}
The comparison of the energies shows that the extremum in Eq.~(\ref{global}) is the global minimum.
Therefore, the ground state at $\overline\nu=1$, i.e., $\nu=-3$ in bilayer graphene, has only one partially filled spin/valley component in the zero-energy Landau band,
and this has a structure identical to the one we obtained in Subsection \ref{spinless}.
The remaining three components are like integer quantum Hall states $\Psi^{(2)}_0$, optimized by the same parameters as discussed in Section \ref{seciqhe}.
Physically, we would lose exchange energy by having several partially filled components, while no gain of correlation energy is possible in the mean-field picture.

At $\nu=-2$, the mean-filled ground state fills the $n=0,1$ orbitals of one spin/valley component completely, while those of the other three components are left empty.
The component with filled zero-energy states is selected by single-particle considerations.
Hund's rule holds in this case.

At $\nu=-1$, the mean-filled ground state fills the $n=0,1$ orbitals of one spin/valley component completely,
while another component is half-filled with the structure discussed in Subsection \ref{spinless}.

At $\nu=0$, the mean-filled ground state fills the $n=0,1$ orbitals of two spin/valley component completely, while those of the other two components are left empty.
Hund's rule holds; additional structure may arise due to the short-range part of the electron-electron interaction, which exhibits the symmerty of the underlying lattice.

As $\tilde E^{(2)}_{1}$ in Eq.~(\ref{spinfulbilayer}) has particle-hole symmetry ($\nu\to-\nu$, $\{\phi_{m\xi\sigma}\}\to\{\phi_{m\xi\sigma}+\pi\}$),
the states at $\nu=1,2,3$ have similar structure.
In particular, Hund's rule holds at $\nu=2$, but not at $\nu=1$ or $\nu=3$.

\section{The quantum Hall ferromagnets in the zero-energy Landau band of ABC trilayer graphene}

In this section we use the notation
\begin{equation}
\overline\nu=\nu+6.
\end{equation}

\subsection{Suppressing spin and valley, $\overline\nu=1$}
\label{fillone}

\begin{table}[htb]
\begin{center}
\begin{tabular}{c|c|c|c|c|c}
\hline\hline
$\left\langle \hat c^\dag_{m'p}\hat c_{mp}\right\rangle$ & $m=0$ & $m=1$ & $m=2$ & $m\ge3$ & $m\le-3$\\
\hline
$m'=0$ & $|z_0|^2$ & $z_0^\ast z_1$ & $z_0^\ast z_2$ & 0 & 0 \\
$m'=1$ & $z_1^\ast z_0$ & $|z_1|^2$ & $z_1^\ast z_2$ & 0 & 0 \\
$m'=2$ & $z_2^\ast z_0$ & $z_2^\ast z_1$ & $|z_2|^2$ & 0 & 0 \\
$m'\ge3$ & 0 & 0 & 0 & $\sin^2\left(\frac{\theta_m}{2}\right)$ &
$\sin\left(\frac{\theta_{|m|}}{2}\right)\cos\left(\frac{\theta_{|m|}}{2}\right)e^{i\phi_{|m|}}$ \\
$m'\le-3$ & 0 & 0 & 0 & $\sin\left(\frac{\theta_m}{2}\right)\cos\left(\frac{\theta_{|m|}}{2}\right)e^{-i\phi_m}$ &
$\cos^2\left(\frac{\theta_m}{2}\right)$ \\
\hline\hline
\end{tabular}
\end{center}
\caption{\label{expabc}
The expectation values relevant to the Hartree-Fock mean-field Hamlitonian of the ABC trilayer using the variational
ground state $\Psi^{(3)}_{1}$ in Eq.~(\ref{variabcone}).
}
\end{table}

\begin{figure}[htbp]
\begin{center}
\includegraphics[width=0.6\columnwidth,keepaspectratio]{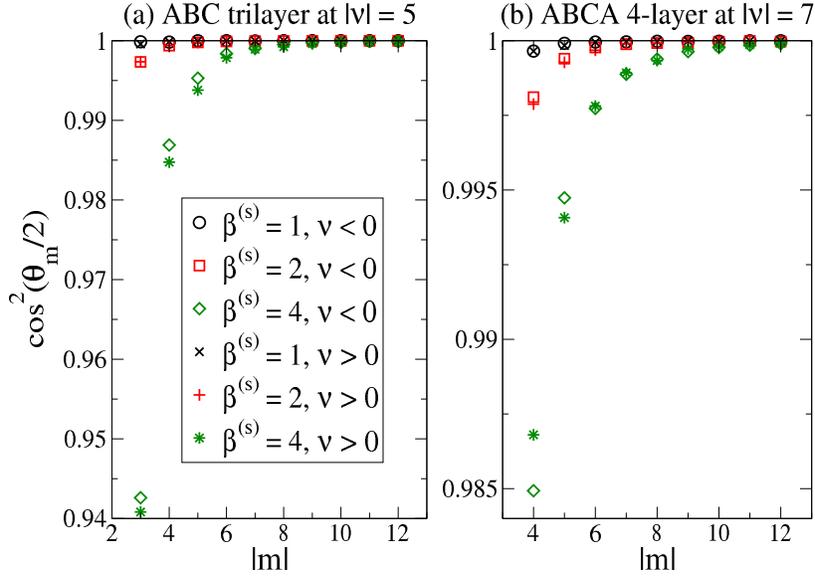}
\end{center}
\caption{\label{one}(Color online)
Projection of the filled orbital to the negative-index Landau orbital at fixed $|m|$ in the quantum Hall ferromagnetic state at
$\overline\nu=1,s-1$ in rhombohedral trilayer and fourlayer graphene.
}
\end{figure}

\begin{figure}[htbp]
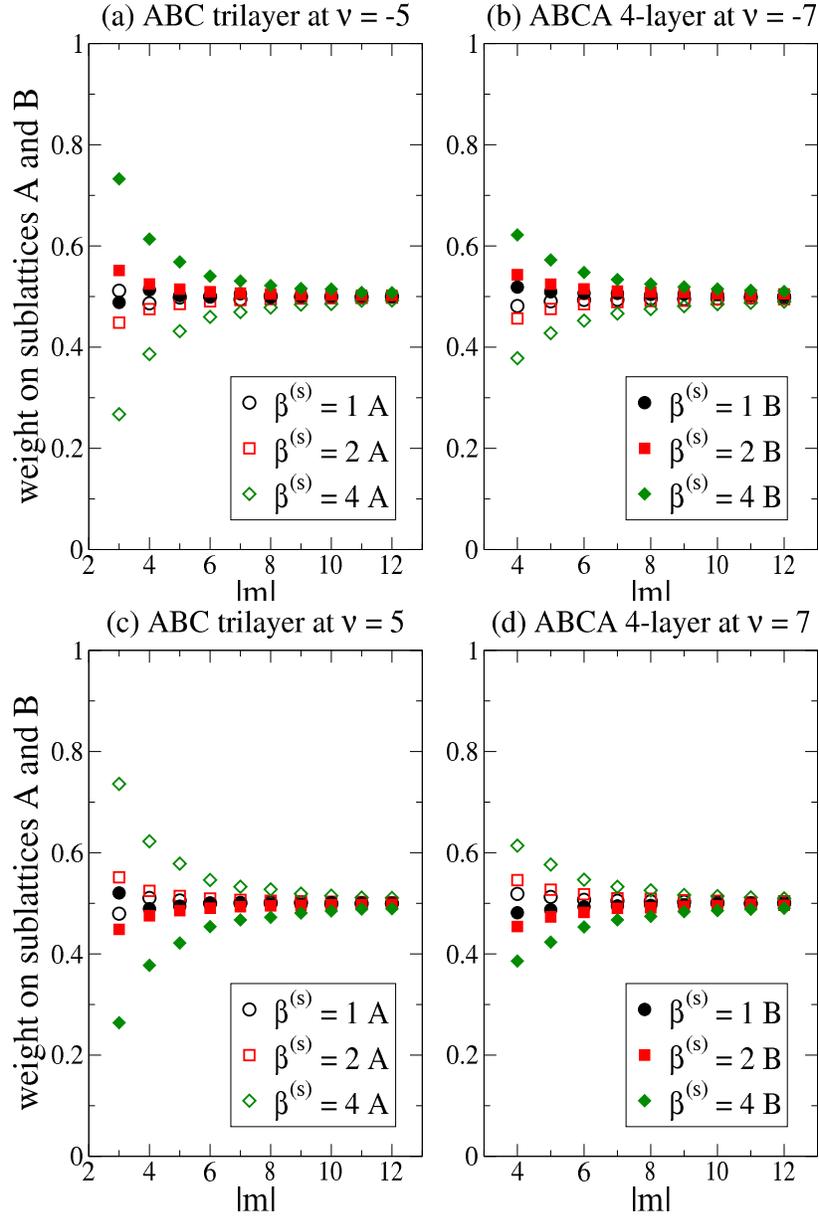

\begin{center}
\includegraphics[width=0.6\columnwidth,keepaspectratio]{oneorbi}
\includegraphics[width=0.6\columnwidth,keepaspectratio]{oneorbi2}
\end{center}
\caption{\label{oneorbi}(Color online)
Weight of the filled orbital in each sublattice at fixed $|m|$ in the quantum Hall ferromagnetic state at $\overline\nu=1,s-1$ in rhombohedral trilayer and fourlayer graphene.
In the upper row $\nu=-2s+1$, i.e., a single Landau level in the zero-energy band is filled, in the lower row $\nu=2s-1$, i.e., the all but one Landau level in the zero-energy band is full.
Particle-hole symmetry changes the preferred sublattice.
}
\end{figure}

We seek the mean-field ground state at relative filling factor $\overline\nu=1$ of the zero-energy Landau band of ABC trilayer graphene
in the form
\begin{equation}
\label{variabcone}
\Psi^{(3)}_{1}=\prod_q
\left(z_0\hat c^\dag_{0q}+z_1\hat c^\dag_{1q}+z_2\hat c^\dag_{2q}
\right)\Psi^{(3)}_0,
\end{equation}
where $\Psi^{(3)}_0$ is the generic form of the integer quantum Hall state at $\nu=-6$ defined in Eq.~(\ref{igheqnsatz}), and
\begin{equation}
\label{normalization}
|z_0|^2+|z_1|^2+|z_2|^2=1.
\end{equation}
The relevant expectation values are shown in Table \ref{expabc}.
The expectation value of the ground state energy per flux quantum in state $\Psi^{(3)}_{1}$ is
\begin{align*}
\frac{E^{(3)}_{1}}{N_\phi}&=
\frac{E^{(3)}_0}{N_\phi}-\frac{1}{2}\int\frac{d^2k}{(2\pi)^2}\frac{e^2}{4\pi\epsilon}\frac{2\pi}{k}
\left[2\sum_{m=3}^\infty\right.\left(
F^{(3)}_{0m}(\mathbf k)F^{(3)}_{m0}(-\mathbf k)|z_0|^2\sin^2\left(\frac{\theta_{m}}{2}\right)+
F^{(3)}_{0,-m}(\mathbf k)F^{(3)}_{-m0}(-\mathbf k)|z_0|^2\cos^2\left(\frac{\theta_{m}}{2}\right)+\right.\\
&F^{(3)}_{0,-m}(\mathbf k)F^{(3)}_{m0}(-\mathbf k)|z_0|^2
\sin\left(\frac{\theta_{m}}{2}\right)\cos\left(\frac{\theta_{m}}{2}\right)e^{i\phi_m}+
F^{(3)}_{0m}(\mathbf k)F^{(3)}_{-m0}(-\mathbf k)|z_0|^2
\sin\left(\frac{\theta_{m}}{2}\right)\cos\left(\frac{\theta_{m}}{2}\right)e^{-i\phi_m}+\\
&F^{(3)}_{1m}(\mathbf k)F^{(3)}_{m1}(-\mathbf k)|z_1|^2\sin^2\left(\frac{\theta_{m}}{2}\right)+
F^{(3)}_{1,-m}(\mathbf k)F^{(3)}_{-m1}(-\mathbf k)|z_1|^2\cos^2\left(\frac{\theta_{m}}{2}\right)+\\
&F^{(3)}_{1,-m}(\mathbf k)F^{(3)}_{m1}(-\mathbf k)|z_1|^2
\sin\left(\frac{\theta_{m}}{2}\right)\cos\left(\frac{\theta_{m}}{2}\right)e^{i\phi_m}+
F^{(3)}_{1m}(\mathbf k)F^{(3)}_{-m1}(-\mathbf k)|z_1|^2
\sin\left(\frac{\theta_{m}}{2}\right)\cos\left(\frac{\theta_{m}}{2}\right)e^{-i\phi_m}+\\
&F^{(3)}_{2m}(\mathbf k)F^{(3)}_{m2}(-\mathbf k)|z_2|^2\sin^2\left(\frac{\theta_{m}}{2}\right)+
F^{(3)}_{2,-m}(\mathbf k)F^{(3)}_{-m2}(-\mathbf k)|z_2|^2\cos^2\left(\frac{\theta_{m}}{2}\right)+\\
&F^{(3)}_{2,-m}(\mathbf k)F^{(3)}_{m2}(-\mathbf k)|z_2|^2
\sin\left(\frac{\theta_{m}}{2}\right)\cos\left(\frac{\theta_{m}}{2}\right)e^{i\phi_m}+
\left.F^{(3)}_{2m}(\mathbf k)F^{(3)}_{-m2}(-\mathbf k)|z_2|^2
\sin\left(\frac{\theta_{m}}{2}\right)\cos\left(\frac{\theta_{m}}{2}\right)e^{-i\phi_m}\right)+\\
&|z_0|^2\left(
F^{(3)}_{00}(\mathbf k)F^{(3)}_{00}(-\mathbf k)|z_0|^2+
F^{(3)}_{01}(\mathbf k)F^{(3)}_{10}(-\mathbf k)|z_1|^2+
F^{(3)}_{02}(\mathbf k)F^{(3)}_{20}(-\mathbf k)|z_2|^2\right)+\\
&|z_1|^2\left(
F^{(3)}_{10}(\mathbf k)F^{(3)}_{01}(-\mathbf k)|z_0|^2+
F^{(3)}_{11}(\mathbf k)F^{(3)}_{11}(-\mathbf k)|z_1|^2+
F^{(3)}_{12}(\mathbf k)F^{(3)}_{21}(-\mathbf k)|z_2|^2\right)+\\
&|z_2|^2\left(
F^{(3)}_{20}(\mathbf k)F^{(3)}_{02}(-\mathbf k)|z_0|^2+
F^{(3)}_{21}(\mathbf k)F^{(3)}_{12}(-\mathbf k)|z_1|^2+
F^{(3)}_{22}(\mathbf k)F^{(3)}_{22}(-\mathbf k)|z_2|^2\right)+\\
&F^{(3)}_{00}(\mathbf k)F^{(3)}_{11}(-\mathbf k)\left(z_0^\ast z_1z_1^\ast z_0+\text{ c.c.}\right)+
F^{(3)}_{11}(\mathbf k)F^{(3)}_{22}(-\mathbf k)\left(z_2^\ast z_1z_1^\ast z_2+\text{ c.c.}\right)+\\
&F^{(3)}_{00}(\mathbf k)F^{(3)}_{22}(-\mathbf k)\left(z_0^\ast z_2z_2^\ast z_0+\text{ c.c.}\right)+\\
&F^{(3)}_{01}(\mathbf k)F^{(3)}_{21}(-\mathbf k)\left(z_0^\ast z_1^2z_2^\ast+\text{ c.c.}\right)+
\left.F^{(3)}_{10}(\mathbf k)F^{(3)}_{12}(-\mathbf k)\left((z_1^\ast)^2z_2z_0+\text{ c.c.}\right)\right]
\end{align*}
By elementary algebra, this simplifies to
\begin{multline}
\label{abcone}
\frac{E^{(3)}_{1}}{N_\phi}=
\frac{E^{(3)}_0}{N_\phi}
-\frac{1}{2}\left(|z_0|^4X_{00}+|z_1|^4X_{11}+|z_2|^4X_{22}\right)-|z_0|^2|z_1|^2(X_{01}+X_{0011})\\
-|z_0|^2|z_2|^2(X_{02}+X_{0022})-|z_2|^2|z_1|^2(X_{12}+X_{1122})-2X_{0121}\Re(z_0^\ast z_1^2z_2^\ast)-\\
-\sum_{m=3}^\infty\left(X_{0m}|z_0|^2+X_{1m}|z_1|^2+X_{2m}|z_2|^2\right)w^+(\theta_m,\phi_m),
\end{multline}
where the exchange integral constants $X_{nm}$ and $X_{nnmm}$ were given Table \ref{xtable}, and
\begin{equation}
X_{0121}=\frac{5\sqrt2}{16}\sqrt{\frac{\pi}{2}}\frac{e^2}{4\pi\epsilon\ell}.\label{x0121}
\end{equation}
The last term in Eq.~(\ref{abcone}) is the exchange interaction of the partially filled zero-energy Landau band and the deeper-lying states.
Further, using Eq.~(\ref{regularization}),
\begin{multline}
\label{abconealt}
\frac{E^{(3)}_{1}}{N_\phi}=
\frac{E^{(3)}_0}{N_\phi}
-\frac{1}{2}\sum_{i=0}^2|z_i|^2(|z_i|^2-1)X_{ii}-2X_{0121}\Re(z_0^\ast z_1^2z_2^\ast)
-\sum_{i<j}\left(\frac{X_{ij}}{2}(2|z_i|^2|z_j|^2-|z_i|^2-|z_j|^2)+X_{iijj}|z_i|^2|z_j|^2\right)-\\
-\frac{1}{2}\sum_{m=3}^\infty\left(X_{0m}|z_0|^2+X_{1m}|z_1|^2+X_{2m}|z_2|^2\right)\sin\theta_{m}\cos\phi_{m},
\end{multline}
In the weak-coupling limit $\beta^{(3)}\to0$, $\theta_m=0$ and the $\phi_m$'s are irrelevant, and the last term in Eq.~(\ref{abconealt}) vanishes.
Only the last-but-one term contains the phases of $z_0$, $z_1$ and $z_2$.
Without loss of generality one can choose $z_1$ as positive real.
Then $z_0=|z_0|e^{i\varphi}$ and $z_2=|\gamma|e^{-i\varphi}$, thus the ground state manifold still has an orbital U(1) symmetry.
Optimizing the magnitudes numerically, we get
\begin{equation}
\label{abczero}
|z_0|^2=0.22418,\quad\quad
|z_1|^2=0.47380,\quad\quad
|z_2|^2=0.30201.
\end{equation}
In general, $E^{(3)}_{1}$ is minimized numerically with a Landau level cutoff at $-M$.
See Figs.~3(a), \ref{one}(a), \ref{oneorbi}(a) for the optimized parameters.
Just like at $\overline\nu=0$, the ground state at $\overline\nu=1$ slightly prefers the sublattice that correspond to the lower component of the spinor.
There is no balanced state, as the particle-hole conjugation always connects quantum Hall ferromagnets at different
filling factor.
Hence the ground state parameters evolve continuously with the interaction strength; no bifurcation occurs.

\subsection{Suppressing spin and valley, $\overline\nu=2$}
\label{filltwo}

We seek the mean-field ground state at relative filling factor $\overline\nu=2$ in the form
\begin{equation}
\label{variabctwo}
\Psi^{(3)}_{2}=\prod_q\left(
\left(z_0\hat c_{0q}+z_1\hat c_{1q}+z_2\hat c_{2q}\right)
\hat c^\dag_{0q}\hat c^\dag_{1q}\hat c^\dag_{2q}
\right)\Psi^{(3)}_0,
\end{equation}
where $\Psi^{(3)}_0$ is the generic form of the integer quantum Hall state at $\nu=-6$ defined in Eq.~(\ref{igheqnsatz}), and normalization holds as in Eq.~(\ref{normalization}).
The relevant expectation values are shown in Table \ref{expabc2}.
Following the same steps as for $\overline\nu=1$, we get
\begin{multline}
\label{abctwo}
\frac{E^{(3)}_{2}}{N_\phi}=
\frac{E^{(3)}_0}{N_\phi}
-\frac{1}{2}\sum_{i=0}^2X_{ii}(1-|z_i|^2)^2-2X_{0121}\Re(z_0^\ast z_1^2z_2^\ast)
-\sum_{i<j}\left(X_{ij}(1-|z_i|^2)(1-|z_j|^2)+X_{iijj}|z_i|^2|z_j|^2\right)-\\
-\sum_{m=3}^\infty\sum_{j=0}^2X_{jm}(1-|z_j|^2)w^+(\theta_m,\phi_m).
\end{multline}
In comparison to the energy expression at $\overline\nu=1$,
\begin{equation}
\frac{E^{(3)}_{2}-E^{(3)}_{1}}{N_\phi}=
\sum_{m=3}^\infty\sum_{j=0}^2X_{jm}|z_j|^2\sin\theta_{m}\cos\phi_{m}
-\frac{1}{2}\sum_{m=3}^\infty\left(X_{0m}+X_{1m}+X_{2m}\right)\sin\theta_{m}\cos\phi_{m}.
\end{equation}
Thus the sign of the last terms in Eqs.~(\ref{emptyenergy}) and (\ref{abconealt}) changes.
Thus if $\{z_0,z_1,z_2,\theta_m,\phi_m\}$ optimizes the ground state energy at $\overline\nu=1$,
$\{z_0,z_1,z_2,\theta_m,\phi_m+\pi\}$ optimizes the ground state energy at $\overline\nu=2$.
Particle-hole symmetry holds, i.e., the missing Landau levels at $\overline\nu=2$ is the same as the filled ones at $\overline\nu=1$.

\begin{table}[htb]
\begin{center}
\begin{tabular}{c|c|c|c}
\hline\hline
$\left\langle \hat c^\dag_{m'p}\hat c_{mp}\right\rangle$ & $m=0$ & $m=1$ & $m=2$ \\
\hline
$m'=0$ & $|z_1|^2+|z_2|^2$ & $z_0z_1^\ast$ & $z_0z_2^\ast$ \\
$m'=1$ & $z_1z_0^\ast$ & $|z_0|^2+|z_2|^2$ & $z_1z_2^\ast$ \\
$m'=2$ & $z_2z_0^\ast$ & $z_2z_1^\ast$ & $|z_0|^2+|z_1|^2$ \\
\hline\hline
\end{tabular}
\end{center}
\caption{\label{expabc2} 
The expectation values relevant to the Hartree-Fock mean-field Hamlitonian of the ABC trilayer using the variational
ground state $\Psi^{(3)}_{2}$ in Eq.~(\ref{variabctwo}).
For the cases of $|m|,|m'|\ge3$, see Table \ref{expabc}.
}
\end{table}

See Figs.~3(a) and \ref{oneorbi}(c) for the optimized parameters.
As compared to $\overline\nu=1$, the preferential sublattices are interchanged.
There is no balanced state, as discussed before.

\subsection{Including spin and valley}

The argument is the same as in Subsection \ref{spin}.
That is, the ground state at $\nu=-3,0,3$ completely fills 1, 2 and 3 spin/valley components, respectively, with $n=0,1,2$ orbitals;
the remaining components have empty $n=0,1,2$ orbital bands.
Hund's rule applies, and the filled components are selected by single-particle considerations.
Here the short-range anisotropic part of the electron-electron interaction also plays a role, as discussed in the
literature.

At $\nu=-5,-2,1,4$ the ground state has one spin/valley component for which the subspace spanned by the $n=0,1,2$ orbitals is 1/3 filled, with a structure
discussed in Subsection \ref{fillone}. The remaining components are either empty or completely filled.

At $\nu=-4,-1,2,5$ the ground state has one spin/valley component for which the subspace spanned by the $n=0,1,2$ orbitals is 2/3 filled, with a structure
discussed in Subsection \ref{filltwo}. The remaining components are either empty of completely filled.
As the partially filled bands have are the particle-hole conjugates of the partially filled bands at $\nu=-5,-2,1,4$,
particle-hole symmetry is obeyed.

\section{The quantum Hall ferromagnets in the zero-energy Landau band of ABCA four-layer graphene}

In this section we use the notation
\begin{equation}
\overline\nu=\nu+8.
\end{equation}

\subsection{Suppressing spin and valley, $\overline\nu=1$}

We seek the mean-field ground state at relative filling factor $\overline\nu=1$ of the zero-energy Landau band of ABCA four-layer graphene
in the form
\begin{equation}
\label{variabcaone}
\Psi^{(4)}_{1}=\prod_q\left(z_0\hat c^\dag_{0q}+z_1\hat c^\dag_{1q}+z_2\hat c^\dag_{2q}+z_3\hat c^\dag_{3q}
\right)\Psi^{(4)}_0,
\end{equation}
where $\Psi^{(4)}_0$ is the generic form of the integer quantum Hall state at $\nu=-8$ defined in Eq.~(\ref{igheqnsatz}), and
\begin{equation}
\label{normfour}
|z_0|^2+|z_1|^2+|z_2|^2+|z_3|^2=1.
\end{equation}
The relevant expectation values are shown in Table \ref{expabca}.

\begin{table}[htb]
\begin{center}
\begin{tabular}{c|c|c|c|c|c|c}
\hline\hline
$\left\langle \hat c^\dag_{m'p}\hat c_{mp}\right\rangle$ & $m=0$ & $m=1$ & $m=2$ & $m=3$ & $m\ge4$ & $m\le-4$\\
\hline
$m'=0$ & $|z_0|^2$ & $z_0^\ast z_1$ & $z_0^\ast z_2$ & $z_0^\ast z_3$ &0&0\\
$m'=1$ & $z_1^\ast z_0$ & $|z_1|^2$ & $z_1^\ast z_2$ & $z_1^\ast z_3$ &0&0\\
$m'=2$ & $z_2^\ast z_0$ & $z_2^\ast z_1$ & $|z_2|^2$ & $z_2^\ast z_3$ &0&0\\
$m'=3$ & $z_3^\ast z_0$ & $z_3^\ast z_1$ & $z_3^\ast z_2$ & $|z_3|^2$ &0&0\\
$m'\ge4$ &0&0&0&0& $\sin^2\left(\frac{\theta_m}{2}\right)$ &
$\sin\left(\frac{\theta_{|m|}}{2}\right)\cos\left(\frac{\theta_{|m|}}{2}\right)e^{i\phi_{|m|}}$ \\
$m'\le-4$ &0&0&0&0& $\sin\left(\frac{\theta_m}{2}\right)\cos\left(\frac{\theta_m}{2}\right)e^{-i\phi_m}$ &
$\cos^2\left(\frac{\theta_{|m|}}{2}\right)$ \\
\hline\hline
\end{tabular}
\end{center}
\caption{\label{expabca} 
The expectation values relevant to the Hartree-Fock mean-field Hamlitonian of the ABCA four-layer using the variational
ground state $\Psi^{(4)}_{1}$ in Eq.~(\ref{variabcaone}).
}
\end{table}

Recall that angular integration in the mean-field Hamiltonian in Eq.~(\ref{meanfield2}) enforces $|n|-|m|=|n'|-|m'|$.
The allowed terms can be classified as follows:
\begin{enumerate}
\item Sixteen cases where $n,n'\in\{0,1,2,3\}$ and $|m|=|m'|\ge4$.
\item Sixteen cases where $m,m'\in\{0,1,2,3\}$ and $|n|=|n'|\ge4$. In fact, these terms are identical to those of Case 1.
\item Sixteen cases $n,n'\in\{0,1,2,3\}$ and $m,m'\in\{0,1,2,3\}$.
\item Fourteen cases enumerated in Table \ref{cases}.
\item Fourteen cases that are identical to those in Table \ref{cases}, with $(n,n')$ and $(m,m')$ interchanged.
\end{enumerate}
\begin{table}[htb]
\begin{center}
\begin{tabular}{c|c|c|c|c}
\hline\hline
$n$ & $n'$ & $m$ & $m'$ & exchange integral\\
\hline
0 & 1 & 0 & 1 & $X_{0011}$\\
1 & 2 & 1 & 2 &$X_{1122}$\\
2 & 3 & 2 & 3 &$X_{2233}$\\
0 & 1 & 1 & 2 &$X_{0121}$\\
1 & 2 & 0 & 1 &$X_{0121}$\\
0 & 1 & 2 & 3 &$X_{0213}$\\
2 & 3 & 0 & 1 &$X_{0213}$\\
1 & 2 & 2 & 3 &$X_{1223}$\\
2 & 3 & 1 & 2 &$X_{1223}$\\
0 & 2 & 0 & 2 &$X_{0022}$\\
1 & 3 & 1 & 3 &$X_{1133}$\\
0 & 2 & 1 & 3 &$X_{0113}$\\
1 & 3 & 0 & 2 &$X_{0113}$\\
0 & 3 & 0 & 3 &$X_{0033}$\\
\hline\hline
\end{tabular}
\end{center}
\caption{\label{cases} 
Terms contributing to the mean-field ground state energy of the state $\Psi^{(4)}_{1}$ of the ABCA four-layer
graphene if each of $n$, $n'$, $m$ and $m'$ are in the zero-energy Landau band, but $n\neq n'$ (hence $m\neq m'$).
We also show the relevant exchange integrals [Eq.~(\ref{exint})], whose values are given in Table \ref{xtable} and
Eqs.~(\ref{x0121}) and (\ref{x1223}).
}
\end{table}
The ground state energy is:
\begin{multline}
\label{abcaone}
\frac{E^{(4)}_{1}}{N_\phi}=
\frac{E^{(4)}_0}{N_\phi}
-\frac{1}{2}\sum_{i=0}^3|z_i|^4X_{ii}-\sum_{i<j}|z_i|^2|z_j|^2(X_{ij}+X_{iijj})-\\
-2X_{0121}\Re(z_0^\ast z_1^2z_2^\ast)-4X_{0213}\Re(z_0^\ast z_1z_2z_3^\ast)
-2X_{1223}\Re(z_1^\ast z_2^2z_3^\ast)
-\sum_{m=4}^\infty\sum_{j=0}^3X_{jm}|z_j|^2w^+(\theta_m,\phi_m),
\end{multline}
where we refer to Table \ref{xtable}, and
\begin{align}
X_{0213}&=\frac{\sqrt3}{16}\sqrt{\frac{\pi}{2}}\frac{e^2}{4\pi\epsilon\ell},\\
X_{1223}&=\frac{5\sqrt3}{32}\sqrt{\frac{\pi}{2}}\frac{e^2}{4\pi\epsilon\ell}.\label{x1223}
\end{align}
Without loss of generality $z_1$ can be taken positive real.
Then the minimization of the three terms in Eq.~(\ref{abcaone}) that depends on the phase of the coefficients
yields only two independent equations; the ground state manifold has U(1) symmetry.
Using Eq.~(\ref{regularization}),
\begin{multline}
\label{abcaonealt}
\frac{E^{(4)}_{1}}{N_\phi}=
\frac{E^{(4)}_0}{N_\phi}
-\frac{1}{2}\sum_{i=0}^3|z_i|^2(|z_i|^2-1)X_{ii}
-\sum_{i<j}\left(\frac{X_{ij}}{2}(2|z_i|^2|z_j|^2-|z_i|^2-|z_j|^2)+X_{iijj}|z_i|^2|z_j|^2\right)-\\
-2X_{0121}\Re(z_0^\ast z_1^2z_2^\ast)-4X_{0213}\Re(z_0^\ast z_1z_2z_3^\ast)
-2X_{1223}\Re(z_1^\ast z_2^2z_3^\ast)
-\frac{1}{2}\sum_{m=4}^\infty\sum_{j=0}^3X_{jm}|z_j|^2\sin\theta_{m}\cos\phi_{m}.
\end{multline}
Notice that only the last term contains and odd number of factors of sine/cosine functions of the phases $\phi_m$.

Keeping the magnitudes $|z_0|$, $|z_1|$, and $|z_2|$ as independent variables and considering the weak coupling
$\beta^{(4)}\to0$ limit, numerical optimization of $E^{(4)}_{1}$ yields
\begin{equation}
\label{abcazero}
|z_0|^2=0.1323,\quad
|z_1|^2=0.3471,\quad
|z_2|^2=0.3549,\quad
|z_3|^2=0.1657.
\end{equation}
For generic $\beta^{(4)}$, a Landau level cutoff $M$ is introduced
by setting $\theta_{m>M}=0$, and the energy minimization is performed numerically.
See Fig.~3(b), \ref{one}(b), and \ref{oneorbi}(b) for the optimized parameters.
Again, there are neither balanced states nor bifurcations, as the particle-hole symmetry connects quantum Hall
ferromagnetic states at different filling factors.

\subsection{Suppressing spin and valley, $\overline\nu=3$}

\begin{table}[htb]
\begin{center}
\begin{tabular}{c|c|c|c|c}
\hline\hline
$\left\langle \hat c^\dag_{m'p}\hat c_{mp}\right\rangle$ & $m=0$ & $m=1$ & $m=2$ & $m=3$ \\
\hline
$m'=0$ & $|z_1|^2+|z_2|^2+|z_3|^2$ & $-z_0z_1^\ast$ & $-z_0z_2^\ast$ & $-z_0z_3^\ast$ \\
$m'=1$ & $-z_1z_0^\ast$ & $|z_0|^2+|z_2|^2+|z_3|^2$ & $-z_1z_2^\ast$ & $-z_1z_3^\ast$ \\
$m'=2$ & $-z_2z_0^\ast$ & $-z_2z_1^\ast$ & $|z_0|^2+|z_1|^2+|z_3|^2$ & $-z_2z_3^\ast$ \\
$m'=3$ & $-z_3z_0^\ast$ & $-z_3z_1^\ast$ & $-z_3z_2^\ast$ & $|z_0|^2+|z_1|^2+|z_2|^2$ \\
\hline\hline
\end{tabular}
\end{center}
\caption{\label{expabca2} 
The expectation values relevant to the Hartree-Fock mean-field Hamlitonian of the ABCA four-layer using the variational
ground state $\Psi^{(4)}_{3}$ in Eq.~(\ref{variabcathree}).
For the cases of $|m|,|m'|\ge4$, see Table \ref{expabca}.
}
\end{table}

We seek the mean-field ground state at relative filling factor $\overline\nu=3$ in the form
\begin{equation}
\label{variabcathree}
\Psi^{(4)}_{3}=\prod_q\left(
\left(z_0\hat c_{0q}+z_1\hat c_{1q}+z_2\hat c_{2q}+z_3\hat c_{3q}\right)
\hat c^\dag_{0q}\hat c^\dag_{1q}\hat c^\dag_{2q}\hat c^\dag_{3q}
\right)\Psi^{(4)}_0,
\end{equation}
where $\Psi^{(4)}_0$ is the generic form of the integer quantum Hall state at $\nu=-8$ defined in Eq.~(\ref{igheqnsatz}), and Eq.~(\ref{normfour}) holds for normalization.
The relevant expectation values are shown in Table \ref{expabca2}, and the ground state energy is
\begin{multline}
\label{abcathree}
\frac{E^{(4)}_{3}}{N_\phi}=
\frac{E^{(4)}_0}{N_\phi}
-\frac{1}{2}\sum_{i=0}^3X_{ii}(1-|z_i|^2)^2-\sum_{i<j}\left(X_{ij}(1-|z_i|^2)(1-|z_j|^2)+X_{iijj}|z_i|^2|z_j|^2\right)-\\
-2X_{0121}\Re(z_0^\ast z_1^2z_2^\ast)-4X_{0213}\Re(z_0^\ast z_1z_2z_3^\ast)
-2X_{1223}\Re(z_1^\ast z_2^2z_3^\ast)
-\sum_{m=4}^\infty\sum_{j=0}^3X_{jm}(1-|z_j|^2)w^+(\theta_m,\phi_m).
\end{multline}
In comparison to the energy expression at $\overline\nu=1$, Eqs.~(\ref{abcaone}) and (\ref{abcaonealt}),
\begin{equation}
\frac{E^{(4)}_{3}-E^{(3)}_{1}}{N_\phi}=
\sum_{m=4}^\infty\sum_{j=0}^3X_{jm}|z_j|^2\sin\theta_{m}\cos\phi_{m}
-\frac{1}{2}\sum_{m=3}^\infty\sum_{j=0}^3X_{jm}\sin\theta_{m}\cos\phi_{m},
\end{equation}
Thus the sign of the last terms in Eqs.~(\ref{emptyenergy}) and (\ref{abcaonealt}) changes.
If $\{z_0,z_1,z_2,z_3,\theta_m,\phi_m\}$ optimizes the ground state energy at $\overline\nu=1$,
$\{z_0,z_1,z_2,z_3,\theta_m,\phi_m+\pi\}$ optimizes the ground state energy at $\overline\nu=3$.
The missing Landau levels at $\overline\nu=3$ is the same as the filled ones at $\overline\nu=1$.
Therefore, particle-hole symmetry holds.
Again, the preferential sublattices at $\overline\nu=1$ and  $\overline\nu=3$ are interchanged;
there are no balanced states or bifurcations.

\subsection{Suppressing spin and valley, $\overline\nu=2$}

We seek the mean-field ground state at $\overline\nu=2$, i.e., when the four-fold orbitally degenerate
zero energy Landau band of rhombohedral four-layer graphene is half-filled, in the following form:
\begin{equation}
\label{variabcatwo}
\Psi^{(4)}_{2}=\prod_q\left(\sum_{i<j}w_{ij}\hat c^\dag_{iq}\hat c^\dag_{jq}\right)\Psi^{(4)}_0,
\end{equation}
where $\Psi^{(4)}_0$ is still the generic state at $\nu=-8$ defined in Eq.~(\ref{igheqnsatz}), and
\begin{equation}
\sum_{i<j}|w_{ij}|=1.
\end{equation}
The relevant expectation values are given in Table \ref{abca2}.

\begin{table}[htb]
\begin{center}
\begin{tabular}{c|c|c|c|c}
\hline\hline
$\left\langle \hat c^\dag_{m'p}\hat c_{mp}\right\rangle$ & $m=0$ & $m=1$ & $m=2$ & $m=3$ \\
\hline
$m'=0$ & $|w_{01}|^2+|w_{02}|^2+|w_{03}|^2$
& $w^\ast_{02}w_{12}+w^\ast_{03}w_{13}$
& $w^\ast_{03}w_{23}-w^\ast_{01}w_{12}$
& $-w^\ast_{01}w_{13}-w^\ast_{02}w_{23}$\\
$m'=1$ & $w_{02}w^\ast_{12}+w_{03}w^\ast_{13}$
& $|w_{01}|^2+|w_{12}|^2+|w_{13}|^2$
& $w^\ast_{01}w_{02}+w^\ast_{13}w_{23}$
& $w^\ast_{01}w_{03}-w^\ast_{12}w_{23}$\\
$m'=2$ & $w_{03}w^\ast_{23}-w_{01}w^\ast_{12}$
& $w_{01}w^\ast_{02}+w_{13}w^\ast_{23}$
& $|w_{02}|^2+|w_{12}|^2+|w_{23}|^2$ 
& $w^\ast_{02}w_{03}+w^\ast_{12}w_{13}$\\
$m'=3$ & $-w_{01}w^\ast_{13}-w_{02}w^\ast_{23}$
& $w_{01}w^\ast_{03}-w_{12}w^\ast_{23}$
& $w_{02}w^\ast_{03}+w_{12}w^\ast_{13}$
& $|w_{03}|^2+|w_{13}|^2+|w_{23}|^2$ \\
\hline\hline
\end{tabular}
\end{center}
\caption{\label{abca2} 
The expectation values relevant to the Hartree-Fock mean-field Hamlitonian of the ABCA four-layer using the variational
ground state $\Psi^{(4)}_{2}$ in Eq.~(\ref{variabcatwo}).
For the cases of $|m|,|m'|\ge4$, see Table \ref{expabca}.
}
\end{table}
The mean-field ground state energy is
\begin{multline}
\label{abcatwo2}
\frac{E^{(4)}_{2}}{N_\phi}=\frac{E^{(4)}_0}{N_\phi}
-\frac{1}{2}\sum_{i=0}^3X_{ii}C_{ii}^2-\sum_{i<j}X_{ij}C_{ii}C_{jj}-\sum_{i<j}X_{iijj}|C_{ij}|^2-\\
-2X_{0121}\Re(C_{21}C_{01})-2X_{0213}\Re(C_{01}C_{32}+C_{31}C_{02})-2X_{1223}\Re(C_{32}C_{12})
-\sum_{m=4}^\infty\sum_{j=0}^3X_{jm}C_{jj}w^+(\theta_m,\phi_m),
\end{multline}
where $C_{ij}$ is the matrix in Table \ref{abca2}.
Notice that only the second row of Eq.~(\ref{abcatwo2}) depends on the phases of $w_{ij}$.
Letting $w_{ij}=|w_{ij}|e^{i\phi_{ij}}$ with $\phi_{ij}$ real,
differentiation of $\frac{E^{(4)}_{2}}{N_\phi}$ by the phases
$\phi_{ij}$ yields only four independent equations.
The dependencies between the six derivatives are as follows:
\begin{gather}
\sum_{i<j}\frac{\partial}{\partial\phi_{ij}}\frac{E^{(4)}_{2}}{N_\phi}=0,\label{dep1}\\
2\frac{\partial}{\partial\phi_{01}}\frac{E^{(4)}_{2}}{N_\phi}+
\frac{\partial}{\partial\phi_{02}}\frac{E^{(4)}_{2}}{N_\phi}-
\frac{\partial}{\partial\phi_{13}}\frac{E^{(4)}_{2}}{N_\phi}-
2\frac{\partial}{\partial\phi_{23}}\frac{E^{(4)}_{2}}{N_\phi}=0.\label{dep2}
\end{gather}
Eq.~(\ref{dep1}) simply expresses the arbitrariness of a global phase,
while Eq.~(\ref{dep2}) states that the ground state energy is unchanged by the following transformation of the phases:
\begin{equation}
\phi_{01}\to\phi_{01}+2\delta,\quad
\phi_{02}\to\phi_{02}+\delta,\quad
\phi_{13}\to\phi_{13}-\delta,\quad
\phi_{23}\to\phi_{23}-2\delta.
\end{equation}
This fact can also be checked by inspecting the middle line of Eq.~(\ref{abcatwo2}).
Therefore, the ground state manifold has U(1) symmetry just like for $\overline\nu=1$ and $\overline\nu=3$.

For convenience, the ground state energy can be written as
\begin{multline}
\label{abcatwoalt}
\frac{E^{(4)}_{2}}{N_\phi}=\frac{E^{(4)}_0}{N_\phi}
-\frac{1}{2}\sum_{i=0}^3X_{ii}C_{ii}(C_{ii}-1)
-\frac{1}{2}\sum_{i<j}X_{ij}(2C_{ii}C_{jj}-C_{ii}-C_{jj})
-\sum_{i<j}X_{iijj}|C_{ij}|^2-\\
-2X_{0121}\Re(C_{21}C_{01})-2X_{0213}\Re(C_{01}C_{32}+C_{31}C_{02})-2X_{1223}\Re(C_{32}C_{12})
-\frac{1}{2}\sum_{m=4}^\infty\sum_{j=0}^3X_{jm}C_{jj}\sin\theta_{m}\cos\phi_{m}.
\end{multline}
Inspecting Table \ref{abca2}, one can check that particle-hole conjugation maps a mean-field ground state
$\{w_{ij},\theta_m,\phi_m\}$ to
$\{w_{01}\leftrightarrow w_{23},w_{02}\leftrightarrow-w_{13},w_{03}\leftrightarrow w_{12},\theta_m,\phi_m+\pi\}$.

In the weak coupling limit $\beta^{(4)}\to0$  the last term in Eq.~(\ref{abcatwoalt}) vanishes,
and the $\theta_m$ angles do not deviate from the value selected by single-particle considerations, i.e., $\theta_m=0$.
Then $E^{(4)}_{2}$ can be minimized numerically in the space of the five independent magnitudes
$\gamma_{01}$, $\gamma_{02}$, $\gamma_{03}$, $\gamma_{12}$, $\gamma_{13}$, and the four
independent phases $\phi_{02}$, $\phi_{03}$, $\phi_{12}$, $\phi_{13}$.
Without loss of generality, one can choose $\phi_{01}=\phi_{23}=0$ and obtain
\begin{equation}
w_{01}=w_{23}=0.311,\quad
w_{02}=0.504 e^{0.035i},\quad
w_{03}=0.387 e^{0.012i},\quad
w_{12}=0.387 e^{-0.014i},\quad
w_{02}=0.504 e^{0.04i}.
\end{equation}

Now, as the band at zero energy is half-filled, particle-hole conjugation in a fixed spin/valley subspace does not
change the filling partial factor $\overline\nu$, there is a chance that ``balanced'' states exist.
These would be analogous to the $|z_0|=|z_1|=\frac{1}{\sqrt2}$ state (or, in the alternative parametrization,
$\theta=\frac{\pi}{2}$) in bilayer graphene at $\overline\nu=1$.
The occupation of the zero-energy orbitals must be such that the last term in Eq.~(\ref{emptyenergy}) is
cancelled, which by Eq.~(\ref{abcatwoalt}) requires $C_{ii}=\frac{1}{2}$.
This is equivalent to $|w_{01}|=|w_{23}|$, $|w_{02}|=|w_{13}|$, and $|w_{03}|=|w_{12}|$.
Then $\theta_m=0$, and the optimization of the $w_{ij}$'s involves six independent parameters only.

As Fig.~2(b) demonstrates, with finite $\beta^{(4)}\gtrsim2.6$, the variational state restricted to the ``balanced'' subspace yields higher energies than the complete search,
which indicates a bifurcation as minima must come in particle-hole conjugate pairs, as only the members of the balanced
subspace are their own particle-hole conjugates.

\section{The evaluation of exchange integrals}
\label{secexchange}

The integral in Eq.~(\ref{exchange}) can be evaluated in closed form as follows.
Assume $n'>n$,
\begin{align}
X_{n'n}&=\int\frac{d^2k}{(2\pi)^2}\frac{e^2}{4\pi\epsilon}\frac{2\pi}{k}|F_{n'n}(\mathbf k)|^2\nonumber\\
&=\frac{e^2}{4\pi\epsilon}\int_0^\infty dk\frac{n!}{(n')!}
\left(\frac{k^2\ell^2}{2}\right)^{n'-n}
\left(L_n^{n'-n}\left(\frac{k^2\ell^2}{2}\right)\right)^2
e^{-k^2\ell^2/2}\nonumber\\
&=\frac{e^2}{4\pi\epsilon\ell}
\int_0^\infty dx\frac{1}{\sqrt2}\frac{n!}{(n')!}x^{n'-n-1/2}
\left(L_n^{n'-n}\left(x\right)\right)^2e^{-x}\nonumber\\
&=
\frac{e^2}{4\pi\epsilon\ell}
\frac{\Gamma(n'-n+1/2)\Gamma(n+1/2)}{\sqrt{2\pi} n! (n'-n)!}
{_3F_2}\left(\frac{1}{2},-n,n'-n+\frac{1}{2}; \frac{1}{2}-n,n'-n+1;1\right).
\end{align}
Here we have introduced $x=\frac{k^2\ell^2}{2}$ and used the identity
\begin{multline}
\int_0^\infty dt\;t^{\alpha-1}e^{-pt}L_m^\lambda(pt)L_n^\beta(pt)=
\frac{p^{-\alpha}\Gamma(\alpha)\Gamma(n-\alpha+\beta+1)\Gamma(m+\lambda+1)}{m!n!\Gamma(1-\alpha-\beta)\Gamma(\lambda+1)}\times\\
\times{_3F_2}\left(-m,\alpha,\alpha-\beta; -n+\alpha-\beta,\lambda+1;1\right)
\end{multline}
with $p=1$, $\lambda=\beta=n'-n$, $m=n$ and $\alpha=n'-n+\frac{1}{2}$.
For $n'<n$, $X_{n'n}=X_{nn'}$ follows by $F^\ast_{n'n}(\mathbf k)=F_{nn'}(-\mathbf k)$.
The same method yields
\begin{multline}
X^{(s)}_{n'n}=\frac{e^2}{4\pi\epsilon\ell}\frac{\Gamma(n'-n+1/2)\Gamma(n+1/2-s)}{\sqrt{2\pi}(n'-n)!}\sqrt{\frac{(n')!}{n!(n-s)!(n'-s)!}}\times\\
\times {_3F_2}\left(\frac{1}{2},-n,n'-n+\frac{1}{2}; \frac{1}{2}+s-n,n'-n+1;1\right).
\end{multline}
$X_{n'n}$ is, of course, just the $s=0$ special case of $X^{(s)}_{n'n}$.

\section{Useful identities}
\label{identities}

Eq.~(\ref{iden}), i.e., $\sum_n|F_{n'n}(\mathbf k)|^2=1$ for arbitrary but fixed $n'$,
is a consequence of the completeness of the Landau orbitals of the two-dimensional electron gas.
Using the orbitals in Eq.~(\ref{etaeq}),
\begin{multline}
\delta(q'-q'')=\int d^2r\eta^\ast_{n'q'}(\mathbf r)\eta_{n'q''}(\mathbf r)
=\int d^2r \int d^2r' \eta^\ast_{n'q'}(\mathbf r)
e^{-i\mathbf k\cdot\mathbf r}\delta(\mathbf r-\mathbf r')e^{i\mathbf k\cdot\mathbf r'}
\eta_{n'q''}(\mathbf r')=\\
=\int d^2r \int d^2r' \eta^\ast_{n'q'}(\mathbf r)e^{-i\mathbf k\cdot\mathbf r}
\left(\sum_{n=0}^\infty\int dq
\eta_{nq}(\mathbf r)
\eta^\ast_{nq}(\mathbf r')\right)
e^{i\mathbf k\cdot\mathbf r'}\eta_{n'q''}(\mathbf r')=\\
=\sum_{n=0}^\infty\int dq
\left(\int d^2r\eta^\ast_{n'q'}(\mathbf r)e^{-i\mathbf k\cdot\mathbf r}\eta_{nq}(\mathbf r)\right)
\left(\int d^2r'\eta^\ast_{nq}(\mathbf r')e^{i\mathbf k\cdot\mathbf r'}\eta_{n'q''}(\mathbf r')\right)=\\
=\sum_{n=0}^\infty\int dq
\delta(q-q'-k_y)F_{n'n}(\mathbf k)e^{-ik_x\ell^2\frac{q+q'}{2}}
\delta(q''-q+k_y)F_{nn'}(-\mathbf k)e^{ik_x\ell^2\frac{q+q''}{2}}=\\
=\sum_{n=0}^\infty|F_{n'n}(\mathbf k)|^2
e^{-ik_x\ell^2\left(\frac{q'}{2} + \frac{k_y+q'}{2}\right)}\delta(q''+k_y-(q'+k_y))
e^{ik_x\ell^2\left(\frac{q''}{2} + \frac{k_y+q'}{2}\right)}
=\sum_{n=0}^\infty|F_{n'n}(\mathbf k)|^2\delta(q''-q').
\end{multline}

Analogous statements hold for the form factors of rhombohedral multilayers:
\begin{equation}
\label{idenmulti}
\sum_n|F^{(s)}_{n'n}(\mathbf k)|^2=1.
\end{equation}
Eq.~(\ref{idenmulti}) is derived from Eq.~(\ref{iden}) as follows.
If $0\le n'<s$,
\begin{equation}
\sum_n|F^{(s)}_{n'n}(\mathbf k)|^2=\sum_{n=0}^{s-1}|F_{n'n}(\mathbf k)|^2+\sum_{|n|\ge s}\frac{1}{2}|F_{n'n}(\mathbf k)|^2=1.
\end{equation}
Otherwise,
\begin{multline}
\sum'_n|F^{(s)}_{n'n}(\mathbf k)|^2=\frac{1}{2}\sum_{n=0}^{s-1}|F_{n'n}(\mathbf k)|^2+\\
+\sum_{n=s}^\infty\frac{1}{4}\left(
|F_{n'n}(\mathbf k)|^2+|F_{n'-s,n-s}(\mathbf k)|^2+F_{n'n}(\mathbf k)F^\ast_{n'-s,n-s}(\mathbf k)+F^\ast_{n'n}(\mathbf k)F_{n'-s,n-s}(\mathbf k)\right)+\\
+\sum_{n=s}^\infty\frac{1}{4}\left(
|F_{n'n}(\mathbf k)|^2+|F_{n'-s,n-s}(\mathbf k)|^2-F_{n'n}(\mathbf k)F^\ast_{n'-s,n-s}(\mathbf k)-F^\ast_{n'n}(\mathbf k)F_{n'-s,n-s}(\mathbf k)\right)=\\
=\frac{1}{2}\sum_{n=0}^\infty|F_{n'n}(\mathbf k)|^2+\frac{1}{2}\sum_{n=s}^\infty|F_{n'-s,n-s}(\mathbf k)|^2=1.
\end{multline}


\begin{thebibliography}{99}
\bibitem{foot1}
The $F_{n'n}(\mathbf q)$ functions are nothing but the single-particle states of the two-dimensional electron gas
in the symmetric gauge, apart from normalization.
They can be shown easily to obey the following orthogonality and completeness relations:
\begin{align}\label{Fortho}
\int d^2q F_{n'n}(\mathbf q)F^\ast_{m'm}(\mathbf q)&=2\pi\delta_{nm}\delta_{n'm'},\\
\sum_{n,n'=0}^\infty F_{n'n}(\mathbf q)F^\ast_{n'n}(\mathbf q')&=\frac{1}{2\pi}\delta(\mathbf q-\mathbf q').
\end{align}
\bibitem{foot2}
One might object that $\hat\rho_{\sigma\sigma'}(\mathbf q)$ is not well-defined in the absence of translational invariance.
Then consider $\hat\rho_{\sigma\sigma'}(\mathbf q)$ as the Fourier transform of
$\hat\rho_{\sigma\sigma'}(\mathbf r,0)$ before the $\epsilon\to+0$ limit is taken.
\end{thebibliography}
\end{document}